\documentclass[12pt,epsf]{article}

\usepackage{graphicx}
\begin{document}
\baselineskip=0.7cm
\newcommand{\EQ}{\begin{equation}}
\newcommand{\EN}{\end{equation}}
\newcommand{\EQA}{\begin{eqnarray}}
\newcommand{\EQN}{\end{eqnarray}}
\newcommand{\EQAN}{\begin{eqnarray*}}
\newcommand{\EQNN}{\end{eqnarray*}}
\newcommand{\e}{{\rm e}}
\newcommand{\Sp}{{\rm Sp}}
\renewcommand{\theequation}{\arabic{section}.\arabic{equation}}
\newcommand{\Tr}{{\rm Tr}}
\newcommand{\lpartial}{\buildrel \leftarrow \over \partial}
\newcommand{\rpartial}{\buildrel \rightarrow \over 
\partial}
\newcommand{\np}{{\rm :}}
\renewcommand{\thesection}{\arabic{section}.}
\renewcommand{\thesubsection}{\arabic{section}.\arabic{subsection}}
\makeatletter
\def\lesim{\mathrel{\mathpalette\gl@align<}}
\def\gtsim{\mathrel{\mathpalette\gl@align>}}
\def\gl@align#1#2{\lower.7ex\vbox{\baselineskip\z@skip\lineskip.2ex%
  \ialign{$\m@th#1\hfil##\hfil$\crcr#2\crcr\sim\crcr}}}
\makeatother
\makeatletter
\def\section{\@startsection{section}{1}{\z@}{-3.5ex plus -1ex minus 
 -.2ex}{2.3ex plus .2ex}{\large}} 
\def\subsection{\@startsection{subsection}{2}{\z@}{-3.25ex plus -1ex minus 
 -.2ex}{1.5ex plus .2ex}{\normalsize\it}}
\makeatother
\makeatletter
\def\lesim{\mathrel{\mathpalette\gl@align<}}
\def\gtsim{\mathrel{\mathpalette\gl@align>}}
\def\gl@align#1#2{\lower.7ex\vbox{\baselineskip\z@skip\lineskip.2ex%
  \ialign{$\m@th#1\hfil##\hfil$\crcr#2\crcr\sim\crcr}}}
\makeatother

\newcommand{\Nt}{\widetilde{N}}
\newcommand{\bra}[1]{\langle #1\vert}
\newcommand{\ket}[1]{\vert #1\rangle}
\newcommand{\braket}[2]{\langle #1\vert #2\rangle}
\newcommand{\bbbk}[4]{{}_1\langle #1|{}_2\langle #2|
                      {}_3\langle #3|#4\rangle_{123}}
\newcommand{\C}{C_{\rm vac}}
\newcommand{\HI}{H_{\rm 3}}
\newcommand{\E}{E}
\renewcommand{\P}{{\bf P}}
\newcommand{\PSV}{P_{SV{\rm -}PS}}

\newcommand{\HSV}{\ket{\HI}_{SV{\rm -}PS}}
\newcommand{\HD}{\ket{\HI}_{DPPRT}}
\newcommand{\HCK}{\ket{\HI}_{CK}}

\newcommand{\vac}{{\rm vac}}
\renewcommand{\v}{{\rm v}}

\renewcommand{\a}{\alpha}
\renewcommand{\b}{\beta}
\newcommand{\inI}{2}
\newcommand{\inII}{3}
\newcommand{\out}{1}
\newcommand{\ainI}{\a_{(\inI)}}
\newcommand{\ainII}{\a_{(\inII)}}
\newcommand{\aout}{\a_{(\out)}}
\newcommand{\ar}{\a_{(r)}}
\newcommand{\as}{\a_{(s)}}
\newcommand{\asp}{\a_{(s')}}
\newcommand{\arp}{\a_{(r')}}
\newcommand{\atp}{\a_{(t')}}
\newcommand{\XI}{X_{\rm I}}
\newcommand{\XII}{X_{\rm II}}
\newcommand{\YI}{Y}
\newcommand{\yI}{Y_{\rm I}^{(1)}}
\newcommand{\YII}{Z}
\newcommand{\aDim}{\Delta^{\langle1\rangle}}
\newcommand{\Ctree}{C^{\langle0\rangle}}
\def\thefootnote{\fnsymbol{footnote}}

\begin{flushright}
hep-th/0510114\\
UT-KOMABA/05-11\\
October 2005
\end{flushright}
\vspace{0.3cm}

\begin{center}
{\Large Extended Fermion Representation \\of Multi-Charge 1/2-BPS Operators in AdS/CFT\\
\vspace{0.2cm}
-- Towards Field Theory of D-Branes --}

\vspace{0.4cm}

Tamiaki {\sc  Yoneya}
\footnote{
e-mail address:\ \ {\tt tam@hep1.c.u-tokyo.ac.jp}}

\vspace{0.3cm}

{\it Institute of Physics, University of Tokyo\\
Komaba, Meguro-ku, Tokyo 153-8902}

\vspace{1cm}
Abstract
\end{center}

We extend the fermion representation of single-charge 1/2-BPS 
operators in the four-dimensional ${\cal N}=4$ super Yang-Mills theory 
to general (multi-charge) 1/2-BPS operators such that  all six directions 
of scalar fields play roles on an equal footing. This enables us 
to construct a field-theorectic representation for a  second-quantized 
system of spherical D3-branes in the 1/2-BPS sector. 
The Fock space of 
D3-branes is characterized by a novel exclusion principle 
(called `Dexclusion' principle), and 
also by a nonlocality which is consistent with the 
spacetime uncertainty relation. 
The Dexclusion principle is realized by 
 composites of two operators, obeying the usual canonical anticommutation relation and 
 the Cuntz algebra, respectively. 
The nonlocality appears as a consequence of  
a superselction rule associated with a symmetry 
which is related to the scale invariance of the super 
Yang-Mills theory. 
 The entropy of the so-called 
superstars,  with multiple charges,  which have been 
proposed to be geometries corresponding to the 
condensation of giant gravitons is discussed 
from our viewpoint and is argued to be consistent with 
the Dexclusion principle. Our construction may be 
regarded as a first step towards a possible new framework 
of general D-brane field theory.

\newpage
\section{Introduction}
Ever since D-branes were recognized as the 
crucial elements of string/M theory, one of difficult 
problems has been the precise description of dynamical creation 
and annihilation of (anti-) D-branes. In order to make real 
progress towards a satisfactory nonperturbative 
formulation of string/M theory, it is desirable to 
develop a new framework in which such dynamical processes 
can be appropriately treated. 

So far, we have two main methods for discussing the 
dynamics of D-branes. One is from the viewpoint 
of open-string theory in which the boundary 
conditions at the open-string ends are 
explicitly taken into account. Various modes of
 open strings are then interpreted 
as describing all the possible dynamical degrees of freedom 
of D-branes. In particular, their transverse positions 
correspond to the lowest massless modes. As such, 
the open-string field theory is a first-quantized 
`configuration-space' formulation of D-branes, even though 
it is second-quantized as the quantum theory of strings. 
The (effective) Yang-Mills theories of D-branes, or matrix models,  are 
in the same vein as open-string field theories. 

In closed-string (field) theories, on the other hand,  D-branes 
are interpreted as a kind of soliton-like excitations.  
A related viewpoint on D-branes has been provided from  
the K-theory classification 
\cite{Witten} of D-branes.  In this case, we 
 start from an appropriate large $N$ limit of D9-brane 
systems, since we can embedd in them an arbitrary 
number of stable or 
unstable (anti) D-branes as solitons or ``lump"  
solutions. Qualitatively similar structure has also been 
appearing in the so-called vacuum string-field 
theory  \cite{Rastelli} of open strings, in which there is no 
propagating degree of freedom for open strings,  
but (unstable) D-branes again 
emerge as soliton-like solutions.

 In all these cases  of the second category, the situation 
 is analogous to discussing the 
creation and annihilation of kinks by using 
sine-Gordon field theory in the case of 
ordinary field theory in two dimensions. 
For kinks, it is well known that the 
system is equivalently described 
by the massive Thirring model in which 
kinks are now regarded as elementary excitations 
corresponding to the Dirac field. The latter is more 
natural and convenient when 
we have to take into account a  large fluctuation 
with respect to pair creation and annihilation of kinks. 
Conversely, the sine-Gordon field theory is 
obtained by the bosonization of the massive 
Thirring model. In the case of D-branes, using this analogy, 
 we do not have the 
 formulation corresponding to the massive Thirring 
model, namely, the field 
theory of D-branes, in which 
D-branes are treated as elementary excitations. 

This is one of the basic motivations of the present work: 
The question we propose to pursue is whether and 
how it is possible to 
second-quantize the open-string or Yang-Mills description 
mentioned above. 
The second quantization in ordinary particle quantum mechanics 
generalizes the Hilbert space of states with fixed number 
of particles by introducing the Fock space representation 
in which all Hilbert spaces with different particle 
number are treated in a unified way.  The single-particle 
wave functions 
are elevated to quantum field 
operators acting upon the Fock space as agents 
creating or annihilating particles. Similarly, we may 
introduce the Fock-type generalized Hilbert 
space and quantum field operators  
creating and annihilating D-branes, by which 
we treat the whole space of super Yang-Mills theories 
with different $N$ in a unified operator formalism. 

At first sight, such an attempt might 
look rather bizarre in view of  expected weirdness 
of field theories for extended objects, except 
when we restrict ourselves to certain very limited 
topological aspects of D-branes.  However, from the 
viewpoint of AdS$_5\times$ S$^5$/MSYM$_4$ correspondence, 
feasibility of such a dynamical structure, at least to some extent,  
is suggested by a recent development on the description of 
1/2-BPS operators on both sides of bulk and boundary 
theories. 

It has been argued that a special class 
of 1/2-BPS operators in U($N$) super Yang-Mills theory,  which are characterized by 
the condition $\Delta (\mbox{conformal dimension})
=J$ with $J$ being the 
angular momentum along a specially chosen 
$U(1)$ direction in S$^5$, are described by 
a matrix model \cite{corley}\cite{Ber1}  with a single (complex) $N\times N$ matrix field 
$Z(t)$ in one dimension, which is reminiscent of the 
old c=1 matrix model. 
This matrix model is equivalently described by 
$N$ free fermions. The second quantization of 
these fermions may bring us something close to the D-brane field, 
since $N$ is nothing but the total D3-brane charge. 

This view has been considerably 
boosted by a more recent 
remarkable work by Lin, Lunin,  and Maldacena \cite{LLM}. 
They showed that the 
supergravity solultions satisfying the same symmetry 
conditions as these 1/2-BPS operators 
are completely classified, under a certain 
smoothness requirement, by a definite boundary condition 
which is formulated on a special two-dimensional plane 
embedded in the bulk. The boundary condition for each 
supergravity solution without singularity 
specifies a configuration which is interpreted as the 
classical phase space configuration corresponding to a 
quantum state of $N$ fermions of the matrix model. 
Quite remarkably, there is a one-to-one holographic 
correspondence between microstates defined  at 
the AdS boundary using fermions and the smooth classical solutions 
in bulk supergravity. 

Actually, as we discuss 
in the next section, the identification of these fermions precisely as 
D3-branes on the gauge-theory side 
suffers from a puzzle which prevents us from 
making such a simple-minded conclusion. 
In the present paper, we propose an entirely new viewpoint for 
 the fermion picture, on deriving it for {\it generic} 1/2-BPS operators 
by treating all six directions of scalar fields $\phi_i$ of 
MSYM$_4$ theory on an equal footing.  
This naturally resolves the puzzle and makes us possible to 
start for a quantum-field theory of D3-branes.  
The construction of field theoretical 
representation of spherical D3-branes 
in the 1/2-BPS sector given below is hopefully 
a small but a first step towards our goal of 
establishing a possible new framework  
for the dynamics of D-branes. 

The paper is organized as follows. In the next section, 
we give a  critical review on the well known 
free-fermion representation of 1/2-BPS operators from 
the viewpoint of possible D-brane field theory. 
In section 3, we first discuss the simple factorization 
property of 2-point functions 
of generic 1/2 BPS operators, which is acturally also 
valid for general 3-point functions and the case of higher-point 
 extremal correlators. 
 It is shown that the emergence of the fermion picture  is 
essentially due to the separation  of the SO(6) degrees of 
freedom from the real dynamical structure which 
contains the dependence of the dynamics on the 
number of D-branes. 
We show that a generalized  version
 of Pauli's exclusion principle 
must be operative, and give 
an operator realization of this D-brane exclusion principle 
(`Dexclusion principle' in short), by introducing composite 
fermion operators using the Cuntz algebra 
as well as the ordinary canonical fermion algebras.  
In section 4, we construct the quantum fields of 
spherical D3-branes whose base spacetime is 
effectively seven dimensional, and discuss the properties of 
their bilinears as creation and annihilation operators 
of giant (and ordinary) gravitons.  We show that 
they reproduce correctly the two-point functions and 
general extremal correlators.   We then discuss the 
nature of nonlocality which is consistent with the spacetime 
uncertainty relation. 
In section 5, we first briefly consider the meaning of 
the Dexclusion principle from the viewpoint of bulk supergravity. 
We then discuss the connection of our 
formulation of mutiple-charged 1/2 BPS 
operators to the so-called 
superstar solutions with mutiple charges on the supergravity side. 
We argue that the entropy of the superstars satisfies 
an inequaltity, 
which is consitent with the Dexclusion principle. 
 The final section is devoted to a summary and concluding 
remarks on future problems. 
In Appendix, we give a derivation of the 
fermion picture in the usual approach, for the purpose of 
setting up our notations and for convenience of the reader 
in comparing our methods with other works.

\section{A critical review: Free fermion as the quantum field of D3-branes? }

A low-energy effective description of (stable) 
 multiple D-branes
is given by matrix fields on their  world volume, 
$ 
X^{\mu}_{ab}
$ 
and their superpartners, 
where $a, b, \ldots$ run from 1 to $N$ with $N$ being the 
number of branes or RR-charge and $\mu$ represents the  
spacetime directions. In particular, 
the transverse positions of D-branes 
are described by the diagonal elements of 
scalar directions ($\mu=i, \, i=1, 2, \ldots, 10-p-1$), 
whereas the off-diagonal elements represent 
the lowest modes of open strings connecting 
different D-branes. 
The theory has the Chan-Paton gauge symmetry 
$
X^{\mu}\rightarrow UX^{\mu}U^{-1}, \, \,  U\in \mbox{U($N$)}
$. 
The gauge symmetry is regarded as the generalization 
of permutation symmetry of multi-particle states 
in the case of ordinary particles, since 
the gauge transformation is reduced to permutation 
of the diagonal elements when the matrix fields are diagonalized. 

The most crucial property of the ordinary second 
quantization is that the 
quantum statistical property 
of multi-particle states is encoded by 
the algebra of field operators. 
Thus the first relevant question towards possible second-quantized 
representation of multi D-brane states is this: Is there any natural 
generalization of the ordinary canonical field algebras  
which corresponds to the gauge symmetry 
as the quantum statistical symmetry 
for D-branes? There is a simple example where we 
already know one possible answer. That is the old $c=1$ matrix model, 
where it is well known that the gauge-invariant Hilbert space 
is equivalent to the Hilbert space of $N$ free fermions. 
Recently, essentially the same logic as for the 
$c=1$ matrix model is argued to be relevant for 
a special class of 1/2-BPS operators in AdS$_5$/MSYM$_4$ 
correspondence, namely in the case of D3-branes. 

The {\it generic} 1/2-BPS operators \cite{nonrenormal} in the theory are 
 \EQ
{\cal O}_{(k_1,k_2, \ldots, k_n)}(x)
\equiv \Big[
{\cal O}_{k_1}(x){\cal O}_{k_2}(x) 
\cdots {\cal O}_{k_n}(x)
\Big]_{(0, k, 0)}
\EN
where 
\EQ
{\cal O}_k(x)\equiv {\rm Tr}
\Big(
\phi_{\{i_1}(x)\phi_{i_2}(x)\cdots \phi_{i_k\}}(x)
\Big)
\EN
is the local product of scalar (hermitian matrix) 
fields $\phi_i \, (i=1, 2, \ldots, 6)$. 
The index $(0, k,0)$, being the standard Dynkin label with 
$k=
\sum_i k_i $, in the right-hand side 
indicates to extract the 
traceless symmetric representation of the SU(4)$\sim$ 
SO(6) R-symmetry group. 
  The dimensions of the representation 
is given by $Dim(k)=(k+1)(k+2)^2(k+3)/12$. 
 The projection to the symmetric traceless representation 
is uniquely possible, just by totally symmetrize the 
tensor indices and subtracting all possible traces. 
There have been given many arguments showing \cite{nonrenormal}
that, in addition to 
conformal dimensions, 2- and 3-point functions of 
the 1/2-BPS operators are not renormalized from 
free-field results. In the case of the so-called extremal 
correlators, the nonrenormalization property is 
argued to be valid for higher-point functions too. 

The usual logic \cite{corley}\cite{Ber1} for the relevance of a free-fermion 
representation is summarized  as follows. First,  let us choose 
a particular plane, say, 5-6 plane,  in the directions transverse 
to D3-branes and consider only the `highest-weight' operators of the form 
\EQ
{\cal O}^J_{(k_1, k_2, \ldots, k_n)}(x)_Z\equiv 
\Tr\Big(Z(x)^{k_1}\Big)\Tr\Big(Z(x)^{k_2}\Big)
\cdots \Tr\Big(Z(x)^{k_n}\Big)
, \quad J=k=\sum_i k_i
\EN
with
\EQ
Z={1\over \sqrt{2}}(\phi_5+i\phi_6).
\EN
The conformal dimensions of these operators 
satisfy $\Delta=J$ with $J$ being the angular 
momentum with respect to the 5-6 plane. 
They form a special class of 1/2-BPS operators. 
In particular, the single-trace operators ($n=1$) 
 essentially correspond to ordinary KK gravitons 
orbiting around the large circle of $S^5$ at the 
intersection with the 5-6 plane. For relatively 
large $J$  and $n$ which are comparable to $N$ 
they have been argued \cite{Balas} to correspond to giant gravitons 
\cite{McG-Aki}, 
spherical D3-branes with dipole-like RR-fields.  
Of course, we can also consider the conjugates 
of these operators with opposite angular momentum 
by replacing $Z$ by $ Z^{\dagger}={1\over \sqrt{2}}(\phi_5-i\phi_6)$. 

Now, suppose we compute two-point functions
\EQ
\langle \overline{{\cal O}}^J_{(\ell_1, \ell_2, \ldots, \ell_m)}(x)_{Z^{\dagger}}
\, {\cal O}^J_{(k_1, k_2, \ldots, k_n)}(y)_Z\rangle. 
\EN
Because of the non-renormalization theorem, 
we can use the (massless) free-field theory with 
action
\EQ
S_{5,6}=-{1\over 2}
\int d^4x \, 
\Tr\Big(
(\partial\phi_5)^2+(\partial \phi_6)^2
\Big)
=-\int d^4x \, \Tr\Big(
\partial Z^{\dagger}\partial Z\Big).
\EN
The two-point functions (with $J=k=\ell$, zero otherwise) 
then take the form
\EQ
\langle \overline{{\cal O}}^J_{(\ell_1, \ell_2, \ldots, \ell_m)}(x)_{Z^{\dagger}}
\, {\cal O}^J_{(k_1, k_2, \ldots, k_n)}(y)_Z\rangle
= f(\{(k), (\ell)\}, N)\, |x-y|^{-2J}, 
\EN
where the function $f(\{(k), (\ell)\}, N)$ is determined by summing over all possible 
combinations of free-field contractions of the 
scalar fields between  
the two operators and hence 
depends  only on $N$ for fixed partitions $\{(k), (\ell)\}$ of the traces 
and $J$. The space-time 
factor $  |x-y|^{-2J}$ comes from the product of $J$ 
free propagators.  Since the function $f(N)$ is 
completely independent 
of the space-time dimensions 4, we can replace the 
4-dimensional free field action by  the
 complex harmonic operator 
in one (Euclidean) dimension, 
\EQ
S^Z=
\int d\tau \,\Tr\Big({d Z^{\dagger}(\tau)\over d\tau}
{d Z(\tau)\over d\tau} + Z^{\dagger}(\tau)Z(\tau)
\Big)
\EN
and simultaneously the two-point functions  
by
\EQ
\langle \overline{{\cal O}}^J_{(\ell_1, \ell_2, \ldots, \ell_m)}(\tau_1)
\, {\cal O}^J_{(k_1, k_2, \ldots, k_n)}(\tau_2)\rangle
=f(\{(k), (\ell)\}, N)\,\e^{-J(\tau_1-\tau_2)}.
\EN
The time parameter $\tau$ can be regarded as 
the radial time of the original 4-dimensional system 
by making identification
\[
\e^{\tau_1-\tau_2}=|x-y|^2, 
\]
which corresponds to a particular 
foliation of  the (Euclidean) 4 dimensional 
base space into $R\times S^3$. Our convention 
for the 4-dimensional base-space metric is Euclidean. 
Thus at least for two-point functions of the above type, 
the 1/2-BPS operators are treated as if they are 
perturbations on D3-branes which are 
uniform along the S$^3$ directions 
of the base space. Essentially we are dealing with only the 
spherical fluctuations of D3-branes. 

Since this matrix model contains only 
one complex matrix $Z$, the above set of two-point functions 
are represented by free fermions, 
using the method which is standard in the theory of 
random matrices. For convenience of 
the reader, we summarize the main steps of the derivation 
in the Appendix.  
In view of the fact that the system with a 
conserved RR-charge of $N$ units is reduced to $N$ free fermions, 
it is tempting to regard them as the fermions 
corresponding to $N$ D3-branes. 
If this identification is correct, the second quantized 
fermion field from the matrix model 
should be interpreted as the quantum 
field of D3-branes.  Is this what we are seeking for? 

In fact, there are some 
puzzling features which do not allow us to 
proceed so straightforwardly. 
The second-quantized fermion 
fields corresponding to the above system are  
defined as 
\EQ
\psi(z, z^*)=\sum_{n=0}b_ np_n(z)\e^{-|z|^2}, 
\EN
\EQ
\psi(z, z^*)^{\dagger}=\sum_{n=0}b_n^{\dagger}
p_n(z^*)\e^{-|z|^2}, 
\EN
where $p_n(z)$ are the normalized monomials,  
\EQ
p_n(z)=  \sqrt{{2^n\over \pi n!}}z^n , \quad 
\int dzdz^* \e^{-2|z|^2}p_n(z)p_m(z^*)=\delta_{nm}
\EN
and $b_n, b_n^{\dagger}$ are fermion creation and 
annihilation operators, 
\[
\{b_n, b_m^{\dagger}\}=\delta_{nm}, \quad 
b_n|0\rangle =0=\langle 0|b_n^{\dagger} .
\]
These field operators satisfy the (lowest Landau level)  
conditions
\EQ
(z+{\partial \over \partial z^*})\psi(z, z^*)=0=
(z^*+{\partial \over \partial z})\psi(z, z^*)^{\dagger}
\EN
reflecting the holomorphic nature of the 
above special set of 1/2 BPS operators. 
The traces of the complex matrix $Z$ are 
represented by fermion bilinears 
\EQ
\Tr\Big(Z^n\Big) 
\leftrightarrow 
\int dzdz^* \psi^{\dagger}(z, z^*)z^n\psi(z, z^*)
={1\over 2^{n/2}}\sum_{q=0}^{\infty}\sqrt{{(n+q)!\over q!}}
b^{\dagger}_{n+q}b_q, 
\EN
\EQ
\Tr\Big((Z^{\dagger})^n\Big) 
\leftrightarrow 
\int dzdz^* \psi^{\dagger}(z, z^*)(z^*)^n\psi(z, z^*)
={1\over 2^{n/2}}\sum_{q=0}^{\infty}\sqrt{{(n+q)!\over q!}}
b^{\dagger}_{q}b_{n+q}.
\EN
In particular, the Hamiltonian is, subtracting the 
zero-point energy,
\[
H=\int dzdz^* \psi(z, z^*)^{\dagger}\Big(
z{\partial \over \partial z}+ zz^*
\Big)\psi(z, z^*)=\sum_{n=0}^{\infty} \, nb_n^{\dagger}b_n.
\]
The term $zz^*$ in the braces is necessary to cancel the contribution 
from the Gaussian part of the wave  function. 
The Euclidean (Heisenberg) equations of motion are 
\EQ
\psi(z, z^*, \tau)=\e^{H\tau}\psi(z, z^*)\e^{-H\tau}, \quad 
\psi^{\dagger}(z, z^*, \tau)\equiv\Big(\psi(z, z^*, -\tau)\Big)^{\dagger}
=\e^{H\tau}\psi^{\dagger}(z, z^*)\e^{-H\tau}. 
\EN

The ground state of $N$ fermions is 
\EQ
|N\rangle \equiv b_{N-1}^{\dagger}b_{N-2}^{\dagger}\cdots 
b_0^{\dagger}|0\rangle.
\EN
The excited states created by acting the above bilinears 
upon $|N\rangle$
are superpositions of various particle-hole pair states 
created in the fermi sea of the ground state. In the language 
of the classification \cite{LLM} of smooth solultions satisfying the same 
symmetry and the energy condition $\Delta=J$ on the 
bulk side, the ground state is nothing but the 
AdS$_5\times S^5$ background itself. 

The puzzle is related to the U(1) R-symmetry 
associated with the angular momentum $J$. 
The symmetry of the fermion system 
associated with the angular momentum is the 
phase transformation
\EQ
z\rightarrow e^{i\theta}z, \quad z^*\rightarrow e^{-i\theta}z^*, 
\label{protation}
\EN
which is equivalently represented in terms of operator 
language, assuming that the fields are scalars with 
respect to SO(6),  as  
$
b_{n}\rightarrow \e^{-in\theta}b_n, \quad 
b_{n}^{\dagger}\rightarrow e^{in\theta}b_n^{\dagger}. 
$
Naively, this would require that even the ground state 
$|N\rangle$ has a nonzero angular momentum 
$J=\sum_{n=0}^{N-1}1=N(N-1)/2$ which is equal to the energy.  
The corresponding bulk theory actually demands that the energy 
and angular momentum should be defined relative to the 
ground state which is the AdS$_5\times$S$^5$ background 
itself. This means that the origin of the 
angular momentum must be redefined 
depending on the choice of state and on the 
number of D3-branes. In the case of angular momentum, such a 
subtraction seems very strange from the viewpoint 
of field  theory of D-branes: ordinarily, a field-theoretical 
angular-momentum operator 
has no ambiguity which would require such a subtraction.

Another puzzle related to the above  is that 
we can choose different U(1) directions and then 
obviously the field operators 
should be regarded as describing different excitation modes of 
D3-branes. On the other hand, even if we are 
treating different excitation modes, the ground state for a given 
$N$ should be one and the same AdS background itself. 
But this is not satisfied at 
least manifestly in the above treatment.  We have to 
identify by hand 
the ground-states defined on different two-dimensional 
planes and hence on the different Hilbert spaces,  
as defining one and the same state. This is very 
unsatisfactory from our viewpoint pointing towards a possible field theory 
of D-branes. 
 
To resolve these puzzles, it is desirable to treat all 
transverse directions of scalar fields on an equal footing. 
Once it could be achieved, the ground state should be manifestly 
SO(6) singlet, and general 1/2-BPS operators with arbitrary 
allowed SO(6) wave functions in the representaion 
$(0, k, 0)$ should arise as independent excited states on it.  
Even apart from the above issues, the possibility 
of extending the fermion description to configurations with 
multiple U(1) charges is an important question by itself, since 
then we would be able to describe the situations where multiple 
giant gravitons are traveling along various different directions 
in $S^5$ simultaneously. The 1/2-BPS condition can still be 
satisfied, corresponding to the 
matrix operators such as  
\EQ
w_{i_1i_2\cdots i_n}\Tr\Big(Z_{i_1}Z_{i_2}\cdots\Big)
\cdots \Tr\Big(\cdots Z_{i_n}\Big)
\EN
where $w_{i_1i_2\cdots i_n}$ is a totally {\it symmetric}  
tensor with respect to $U(3)$ group and the 
scalar matrices are arranged into 
the complex basis as 
\[
Z_1={1\over \sqrt{2}}(\phi_1+i\phi_2), \quad 
Z_2={1\over \sqrt{2}}(\phi_3+i\phi_4), \quad 
Z_3={1\over \sqrt{2}}(\phi_5+i\phi_6), 
\]
corresponding to  three Cartan directions of SO(6). 
Note that, if the total symmetrization of $w$-tensors were not imposed and 
hence other SO(6) representations than $(0, n, 0)$ were 
assumed, they could 
lead to lower BPS operators with 1/4 or 1/8 supersymmetries. 
Note that even this set of operators does not 
exhaust the whole set of 1/2-BPS operators discussed above, 
since we could still include the conjugates of these complex 
matrices by 
explicitly taking into account the traceless condition with respect to 
SO(6) indices.\footnote{For a simple example, see 
the expression (\ref{zzbarsinglet}) in the concluding section.}
From our point of view, it does not seem natural to 
decouple, as suggested in \cite{Ber1}, other directions than the 
5-6 plane by introducing an artificial parameter which 
violates SO(6) symmetry. Of course, it is always possible to 
introduce symmetry breaking terms for the purpose of 
studying the system with reduced degrees of freedom\footnote{
For such a study, though in an entirely different context, 
see a recent work \cite{LM}, appearing in the course of
completing the present paper,  which contains a 
discussion on a Witten index counting 1/2-BPS states in theories with 
 supersymmetry SU($2|4$). }
after constructing a fully symmetric field theory.

\section{A generalized exclusion principle and composite fermions}
\setcounter{equation}{0}

In order to generalize the fermion picture to generic 
1/2-BPS operators, it seems at first sight that we have to 
invent some appropriate generalization of diagonalization 
technique to the case of many matrices. This 
has long been one of difficult unsolved problems in matrix 
field theories. Fortunately, however, in the present case of 1/2-BPS 
operators, we can justifiably use the free-field approximation 
for the matrix fields because of various non-renormalization 
properties \cite{nonrenormal}. We can compute 
{\it exact} correlation functions explicitly at least for two or 
three-point functions (and some special classes of 
higher-point functions, such as the so-called 
extremal correlators) and examine directly 
whether they allow any interpretation  
by some generalized fermion picture. We will  indeed argue 
 that the essence of the emergence of the fermion 
picture is {\it not} in the choice of a special plane. 

\subsection{Factorization theorem and the separation of degrees of freedom}

Let us start from considering the 
properties of correlation functions for generic 1/2-BPS 
operators.  
We choose the basis for general 1/2-BPS operators 
in the form
\EQ
{\cal O}^I_{(k_1, k_2, \ldots, k_n)}(x)
\equiv  w^I_{i_1\cdots i_k}
\Tr\Big(\phi_{i_1}\cdots \phi_{i_{k_1}}\Big)
\cdots
\Tr\Big(\phi_{i_{k-k_n+1}}\cdots\phi_{i_k}\Big), 
\EN
where $k=k_1+k_2+\cdots k_n$, with 
$(k_1, \ldots, k_n)$ being the number of matrices in 
 partitioning them 
into multi-traces,  is equal to the 
conformal dimension $\Delta=k$. Though not orthogonal 
in the sense of conformal operators, 
this basis is most convenient for our 
purpose in this section.  If we wish, we can go to the 
orthogonal basis using the Schur polynomial method as 
in \cite{corley}. For the SO(6) part, 
$\{w^I_{i_1\cdots i_k}\}$ denotes the basis for 
totally symmetric traceless tensors. When necessary,  one can 
arrage them so that they satisfy the orthonormality condition
$
\langle w^{I_1} w^{I_2}\rangle\equiv 
w^{I_1}_{i_1\cdots i_k}w^{I_2}_{i_1\cdots i_k}=\delta^{I_1I_2}, 
$
defining the S$^{5}$ harmonics 
$
Y^I[\phi_i]=
w^I_{i_1\cdots i_n}\phi_{i_1}\cdots \phi_{i_n}. 
$
However, the orthogonality is not necessary for our present 
purpose. 

Now we consider two-point functions using 
free-field theory. As before, we can actually replace 
the base space by 
a one-dimensional space whose coordinate is identified 
with the radial time ($\tau$) of the original 4 dimensional Euclidean 
world volume, 
\[
\langle {\cal O}^{I_1}_{(k_1, k_2, \ldots, k_n)}(\tau_1)
{\cal O}^{I_2}_{(\ell_1, \ell_2, \ldots, \ell_n)}(\tau_2)\rangle .
\]
Because of the traceless condition, the correlator 
is expressed by all possible contractions 
of  matrix fields between the two {\it separate} sets of the 
products of traces of matrix fields at $\tau_1$ and $\tau_2$. 
A contraction gives the Kronecker delta for  the SO(6) indices times 
the factor $e^{-(\tau_1-\tau_2)}/2$. Because of the total 
symmetry of the tensors $w^{I_1}$ and  $w^{I_2}$, 
all the different SO(6) contractions  associated with these 
contractions always give one and the same factor 
$\langle w^I w^J\rangle$. 
Thus the net result takes a 
factorized form 
\EQ
\langle {\cal O}^{I_1}_{(k_1, k_2, \ldots, k_n)}(\tau_1)
{\cal O}^{I_2}_{(\ell_1, \ell_2, \ldots, \ell_n)}(\tau_2)\rangle
=\langle w^{I_1} w^{I_2}\rangle G(\{(k), (\ell)\}, N)e^{-k(\tau_1-\tau_2)}, 
\EN 
where the factor $G(\{(k), (\ell)\}, N)$, which 
is in fact identical with the previous $f(\{(k), (\ell)\}, N)$, 
 is independent of the SO(6)
tensor wave functions and of the coordinate $\tau$. 
Of course, the factorization of SO(6) invariant 
$\langle w^{I_1}w^{I_2}\rangle$ itself is just a consequence of the SO(6) symmetry 
(Wigner-Eckert theorem) which is valid for any theory 
satisfying a global SO(6) symmetry. What is special to the 
1/2 BPS operators for ${\cal N}=4$ SYM$_4$ is that the remaining factor  
can be replaced by  the sum of all possible contractions  
of free-field theory of a {\it single hermitian} matrix field. 
This is valid universally for all partitioning for the matrix traces. 
Namely, 
we have 
\EQ
G(\{(k), (\ell)\}, N)\e^{-r(\tau_1-\tau_2)}
=\langle 
{\rm :}{\cal O}^r_{(k_1, \ldots, k_n)}(\tau_1)_M{\rm :}\, 
{\rm :}{\cal O}^r_{(\ell_1, \ldots, \ell_n)}(\tau_2)_M{\rm :}\rangle_M
\label{factorizationtheorem}
\EN
where the matrix operators are 
defined in the same way as before by replacing 
the hermitian matrices $\phi_i$ by the single 
{\it hermitian} matrix field $M(\tau)$;
\EQ
{\cal O}^r_{(k_1, \ldots, k_n)}(\tau_1)_M
 \equiv 
\Tr\Big(M^{k_1}\Big)\Tr\Big(M^{k_2}\Big)
\cdots \Tr\Big(M^{k_n}\Big)
\EN
with the action $S_M=-{1\over 2}\int d\tau \Tr(\dot{M}^2 +M^2)$. 
The normal product symbol $\np \cdots \np$ indicates that 
 no contraction is allowed inside. For the validity of this 
reduction into the single-matrix model, it is sufficient, because of the 
free-field approximation, that the 
SO(6) factors for arbitrary contractions always give 
a uniquely fixed quantity. This is satisfied  also in the case of
 3-point functions and 
 for general {\it extremal} $n$-point functions. 
 In the present section, however, 
we restrict ourselves only to two-point functions.

What we should learn from these almost 
trivially looking observations is that the emergence of the one-matrix 
model for describing 1/2 BPS operators is 
essentially due to the separation of degrees 
of freedom into the purely matrix degrees of freedom 
and the purely kinematical SO(6) degrees of freedom. 
This is a {\it dynamical} property which can not be 
explained by the SO(6) symmetry alone, and whose 
origin lies in non-renormalization of the 1/2-BPS 
correlators.\footnote{For example, 
the Wigner-Eckert theorem alone cannot say anything about the relations 
of invarints, after $\langle w^{I_1}w^{I_2}\rangle$ being factored out, 
for various {\it different} configurations of the partitions of matrix traces. }
 The choice of a special plane as 
in the usual argument is not necessary. The usual 
manipulation, reviewed in the Appendix, is perfectly 
valid. We are, however, saying that the one-matrix model 
can play key roles even if we treat all SO(6) directions 
equivalently. As we see below, this viewpoint provides a  natural 
resolution of our puzzles. 

Actually, it is also possible and is more convenient to 
use a single {\it complex} matrix field $Z$, instead of the 
hermitian matrix $M$, since then we can automatically 
avoid the normal ordering prescription in the case 
of two-point functions (and also for 
general extremal correlators). This simply amounts to using a 
coherent-state representation for one-dimensional (matrix) 
harmonic oscillator. 
Thus we have a simple result
\newpage
\EQ
\langle {\cal O}^{I_1}_{(k_1, k_2, \ldots, k_n)}(\tau_1)
{\cal O}^{I_2}_{(\ell_1, \ell_2, \ldots, \ell_n)}(\tau_2)\rangle
=\langle w^{I_1}w^{I_2}\rangle 
\langle \overline{{\cal O}}^k_{(k_1, k_2, \ldots, k_n)}(\tau_1)_{Z^{\dagger}}
{\cal O}^k_{(\ell_1, \ell_2, \ldots, \ell_n)}(\tau_2)_Z\rangle
\label{factorcomplexrep}
\EN
using the  same complex matrix model as for the 
special 1/2-BPS operators satisfying $\Delta=J=k$. 
If we wish, we can further replace this complex matrix model by the 
first-order model with the action, 
\[
S=2\int d\tau \, \Tr\Big[-
Z^{\dagger}{\partial \over \partial \tau}Z +Z^{\dagger}Z
\Big]
\]
instead of  the second order action. 
The difference between the first and second order 
models lies only in Green functions. In the former, 
the propagator is 
\[
\langle Z^{\dagger}_{ab}(\tau_1)\, Z_{cd}(\tau_2) 
\rangle =\cases{
{1\over 2}\e^{-(\tau_1-\tau_2)}\delta_{ad}\delta_{bc} & $\tau_1>
\tau_2$
\cr 
0 & $\tau_1<\tau_2$ \cr
}.
\]
Remember that in the case of the previous second order action, 
the propagator is \\
$\langle Z^{\dagger}_{ab}(\tau_1)\, Z_{cd}(\tau_2) 
\rangle \propto \exp(-|\tau_1-\tau_2|)$. 
For two-point functions and higher-point extremal correlators, 
this difference does not 
matter, since by our definition the operators with 
$Z^{\dagger}$'s always appear after those with $Z$'s.

We emphasize that, though we are using the same 
notation, the meaning of the complex matrix $Z$  
is now entirely different from the previous case with the 
special operators satisfying $\Delta=J$ with a single U(1) charge $J$. 
Here it is introduced 
merely as a technical device 
(coherent-state representation) in order to avoid 
contractions automatically inside each single matrix operators 
at a given time. 
The origin of the lowest Landau level condition is 
nothing other than the normal ordering condition in (\ref{factorizationtheorem}). 
Therefore, its phase transformation is nothing to do 
with the angular momentum of a particular SO(2) subgroup 
of the R-symmetry. Thus, in our formulation, the matrix degrees of 
freedom are completely inert under SO(6) from the outset. 
The reason why we were led to interpret the phase 
rotation (\ref{protation}), in the ordinary derivation, 
 as the SO(2) rotation as the 
subgroup of SO(6) is a coincidence of their charges  between 
the SO(6) wave functions $w^I$'s and the wave functions 
of matrix model: it occurs 
when we use the complex bases {\it both} for SO(2) $\subset$ SO(6) 
and the matrix model. This explains why we had to 
subtract the charges of the ground state when 
we reinterpret the phase rotation by the 
fermion creation-annihilation operators.

Now, our problem is how to interpret 
these two-point functions by treating generic 
1/2-BPS operators with different SO(6) 
wave functions as {\it independent} excitation modes 
of the would-be generalized fermionic fields of D3-branes.    
Under the motivations explained in the 
previous section, our task is therefore to 
\begin{enumerate}
\item[(1)] define the D-brane fields 
which involve coordinate dependence 
corresponding to all the scalar directions $\phi_i$ on an equal footing; 
\item[(2)] interpret the generic 1/2-BPS matrix 
operators as bilinear functions (and their products) in terms 
of D-brane fields,  
\end{enumerate}
 such that the above simple factorized form (\ref{factorcomplexrep}) 
 of two-point correlation fucntions is obtained. 
 
In the fermion picture, the ground state $|N\rangle$ of $N$ D3-branes 
is the lowest-energy state which is occupied by $N$ fermions. 
We assume that the ground 
states with increasing $N$ are 
consecutively constructed from ground states  
with smaller number of fermions by acting creation operators of 
D3-brane. All these ground states with different $N$ must be
 SO(6) singlet. Then the corresponding 
creation operators must also be SO(6) singlet. 
We denote such SO(6) singlet fermion creation 
operators by $b_{n, 0}^{\dagger}$ where $n$ labels 
the energy levels and $0$ designates that they are 
SO(6) singlet. 
Thus
\EQ
|N\rangle =
b^{\dagger}_{N-1, 0}b^{\dagger}_{N-2, 0}\cdots 
b^{\dagger}_{0,0}|0\rangle. 
\EN
The fermionic nature requires 
\EQ
\{b^{\dagger}_{n, 0}, b^{\dagger}_{m, 0}\}=0
\EN
for arbitrary pair of energy levels $(m, n)$. 
We also introduce the conjugate operators and states
\EQ
\langle N| =\langle 0|
b_{0, 0}\cdots b_{N-2,0} 
b_{N-1,0}, 
\EN
\EQ
\{b_{n, 0}, b_{m, 0}\}=0, 
\EN
satisfying 
\EQ
\langle N'|N\rangle =\delta_{N'N}. 
\EN
Ordinarily, the orthogonality condition is ensured by the 
standard anticommutation relation of creation and 
annihilation operators $\{b_{n, 0}, b^{\dagger}_{m, 0}\}=
\delta_{nm}$ with the vacuum conditions 
\EQ
b_{n, 0}|0\rangle =0=\langle 0|b_{m, 0}^{\dagger}. 
\EN

For our purpose here, however, 
it is important to keep in mind that 
the anticommution relation between the creation 
and annhilation operators are not completely 
compulsory for ensuring the orthogonality. 
It is sufficient to assume that the annihilation operator 
indeed annihilates one fermion in the usual manner 
{\it when acting upon the gound state}, 
\EQ
b_{n, 0}|N\rangle =\cases{
(-1)^{N-1-n}\, b^{\dagger}_{N-1, 0}b^{\dagger}_{N-2, 0}\cdots 
\widehat{b^{\dagger}_{n,0}}\cdots b^{\dagger}_{0,0}|0\rangle & for  $n<N$ \cr
 0 & otherwise \cr}
 \label{fermionicnature}
\EN
where the object below the hat is absent.  
We do not assume the operator 
anticommutation relation $\{b_{n, 0}, b^{\dagger}_{m, 0}\}=
\delta_{nm}$, and will shortly see that there is still a consistent 
operator algebra which satisfies the requirement (\ref{fermionicnature}) 
and orthogonality condition simultaneously. 

\subsection{`D'-exclusion principle and Cuntz algebra -- a generalized Pauli principle 
for D-branes}
\begin{center}
\begin{figure}
\begin{picture}(180,160)
\put(100,-10){
\includegraphics[width=250pt]{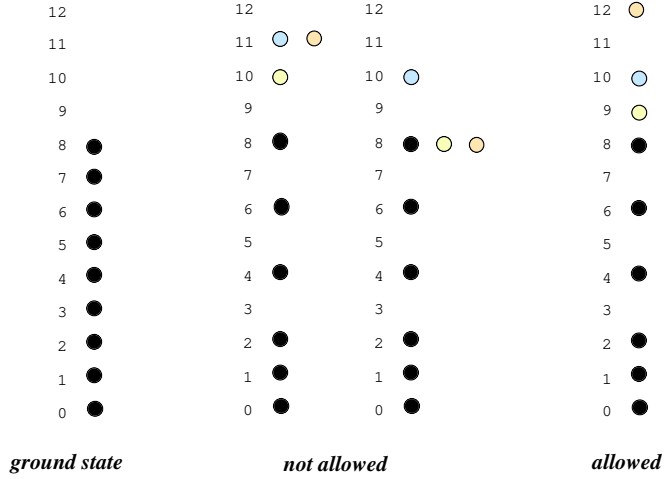}}
\end{picture}
\caption{This illustrates the Dexclusion principle. The vertical axis 
indicates the energy levels. The dots are 
occupied energy levels: black=SO(6) singlet, color(gray) =
nonsinglet. The same energy levels cannot be occupied by two or more 
particles simultaneously, 
irrespectively of their SO(6) states. }
\label{fig1}
\end{figure}
\end{center}

Next we have to consider excited states. From now on, 
we assume for definiteness that the SO(6) basis $w^I$ is orthonormalized, 
$\langle w^{I_1}w^{I_2}\rangle =\delta^{I_1I_2}$. 
As the first step, let us consider an 
excited state which corresponds to 
the action of the single trace operator 
\[
{\cal O}^I_{(k)}=w^I_{i_1, i_2, \ldots, i_k}\Tr\Big(
\phi_{i_1}\phi_{i_2}\cdots \phi_{i_k}\Big).
\]
In the fermion picture, this must correspond to creating a pair 
of particle and hole in the fermi sea 
of the ground state in such a way that  the 
created pair has the designated SO(6) state 
$I$ and lifts the energy by $k$ units 
(=conformal dimension). 
The annihilated single-particle state must be one (energy=$n$) 
of the singlet states occupied in $|N\rangle $ and 
the created state must then have energy $n+k$ and 
carry the nontrivial SO(6) index $I$. 
The two-point function of this 
operator with its conjugate has to satisfy the above 
factorization property (\ref{factorcomplexrep}). Namely, it must have the 
 $N$ dependence which is identical with the case of the 
 special 1/2-BPS 
operator with $\Delta=J$, apart from the 
SO(6)  factor. Denoting the creation and annihilation 
operators with non-singlet index $I$ 
by $b^{\dagger}_{n, I}$ and $b_{n, I}$, 
the relevant part of the corresponding 
fermion bilinear (its conjugate)  
would then take the following form
\[
2^{-k/2}\sqrt{{(n+k)!\over n!}}\, b^{\dagger}_{n+k, I}b_{n, 0}, 
\quad 2^{-\ell/2}\sqrt{{(n+\ell)!\over n!}}
b^{\dagger}_{n, 0} b_{n+\ell, I'}
\]
with the orthogonality condition
\EQ
\langle N| 
b^{\dagger}_{n, 0} b_{n+\ell, I'}
\, b^{\dagger}_{n+k, I}b_{n, 0}|N\rangle
= \cases{\delta_{k, \ell}\delta_{I, I'} & for 
$k+n\ge N, n<N$\cr
0 & otherwise\cr} .
\EN
The vanishing condition of the second line is one of the crucial 
and inevitable requirements in our formulation, 
originating from the factorization property. 
This means in particular that any two 
single-particle D3-brane states with the same energy levels 
must be mutually exclusive even if they have 
different SO(6) indices, 
\EQ
b^{\dagger}_{n+k, I}b^{\dagger}_{N-1, 0}b^{\dagger}_{N-2, 0}\cdots 
\widehat{b^{\dagger}_{n, 0}}\cdots 
b^{\dagger}_{0,0}|0\rangle=0 \quad \mbox{for} \quad n+k <N , \, 
k\ne 0
\quad \mbox{with aribitrary $I$} . 
\EN
Thus, we are now 
encountering a stronger version of Pauli's exclusion principle, 
which should perhaps be interpreted as a signal 
of the strange quantum statistical property of D-branes 
represented by gauge symmetry. 
If the particle with nontrivial SO(6) representation 
obeys the ordinary Pauli principle that only excludes the 
case of occupying the same state with completely 
the same labels with respect to all quantum numbers, the states of 
particle-hole pairs would have higher 
degeneracy, since created particles can have energies which are 
 lower than the fermi surface of the ground state. 
Of course, in this special case 
of single-trace operators, it is sufficient to assume the above 
exclusiveness between the singlet and non-singlet 
representations. We will later see that, to treat multi-trace 
operators consistently with (\ref{factorcomplexrep}), 
we have to generalize this exclusion 
property to the cases between arbitrary two SO(6) wave functions, irrespectively of SO(6) representations. 
We propose to call this generalized exclusion principle for D3-branes, 
{\it D-exclusion} (or `Dexclusion') principle (DEP). 
See Fig. \ref{fig1}. 
\footnote{
This should not be confused with the so-called {\it stringy} exclusion 
principle \cite{stro}. }  

\vspace{-1cm}
It is not possible to realize the DEP by the usual fermionic algebra. 
A natural possibility suggested from the 
separation of degrees of freedom  is to assume that 
the creation and annihilation operators are composites 
of two independent operators acting in different 
spaces, each of which 
carries either the SO(6) vector labels or the energy labels 
separately as 
\EQ
b^{\dagger}_{n, I}=c^{\dagger}_I  \, \otimes b_{n}^{\dagger}, \,  \quad 
b_{n,I}=c_I\,\otimes b_{n}. 
\EN
We call the $b$-type operators  `energy' operators and the 
$c$-type `vector' operators. Energies and particle numbers are 
carried only by the energy operators. For brevity of notation, 
the product symbol `$\otimes$' will be suppressed below. 
Since these two kinds of operators are assumed to be mutually commutative, 
$c^{\dagger}_I\, b_{n}^{\dagger}=b_{n}^{\dagger}\, c^{\dagger}_I, \, 
c_Ib_m^{\dagger}=b^{\dagger}_mc_I, etc $, 
 the DEP is satisfied 
\EQ
b^{\dagger}_{n, I}b^{\dagger}_{n, I'}=0, 
\EN 
 providing that the 
energy creation-annihilation operators satisfy the 
standard canonical fermion algebra, 
\EQ
\{b_n, b^{\dagger}_m\}=\delta_{n,m}, \quad 
\{b_n, b_m\}=0=\{b_n^{\dagger}, b_m^{\dagger}\}
\EN
and the associated vacuum condition 
$b_n|0\rangle=0=\langle 0|b^{\dagger}_n$. 
The energy is carried only by the energy operators, and 
the Hamiltonian is simply 
\EQ
H=\sum_{n=0}^{\infty}nb_n^{\dagger}b_n, 
\EN
resulting the  (Euclidean) time dependence of the composite 
operators as $c_Ib_{n}\e^{-n\tau}=c_Ib_n(\tau)$ and 
$c_I^{\dagger}b_{n}^{\dagger}\e^{n\tau}=c_Ib_n^{\dagger}
(\tau)$.\footnote{
The composite fermion operators defined here have some 
formal resemblance to  composite operators 
which are introduced \cite{jain} to describe fractional 
quantum Hall effect.  Though the origin of compositeness is entirely 
different, it would be interesting if the analogy could be 
formulated in a more physical way. A different aspect 
of analogies with fractional quantum Hall effect which may be related to 
our discussion in section 5 has recently been considered in  
\cite{Dai}}

Actually, there arises an immediate problem in this proposal. 
Once we introduce the composite operators as above, there is a
danger that 
the ground states (for $N\ge 2$) (and hence any 
states) could be infinitely degenerate, 
even if we require that they are SO(6) singlet. 
The reason is that there are infinitely many 
different ways of combining vector creation operators $c^{\dagger}_I$'s 
into the singlet representation without any cost of energy. 
It seems that, at the price of realizing the DEP to reduce the degeneracy 
of excited states, we  are facing  the problem of 
infinite degeneracy of the ground state, and hence an 
infinitely degenerate Hilbert space. As it stands, this 
system may not be acceptable as a sensible physical system. 

This difficulty can be saved if we can interpret the  
degeneracy as being due to the existence of 
an infinite number of superselection sectors. 
 Let us  assign charges (that we call `S'-charges) to 
these operators by the following phase transformations 
\EQ
b^{\dagger}_{n}\rightarrow \e^{in\theta}b^{\dagger}, 
\quad c^{\dagger}_I\rightarrow \e^{-ik(I)}c^{\dagger}_I
\EN
(similarly for their conjugate operators) where
$k(I)$ is the level (=the rank of the traceless 
symmetric tensors) of the SO(6) state $I$. 
The original creation (or annihilation) operators of D3-brane 
have the charges $n-k(I) \, (\mbox{or} \, \, -n+k(I))$, 
\[
b_{n, I}^{\dagger}\rightarrow \e^{i(n-k(I))\theta}b_{n,I}^{\dagger}. 
\]
Thus, an excitation of  SO(6) representation of higher rank lowers the 
S-charges. 
We assume that the multi-particle states must have a definite 
S charge for a given $N$. 
The S charge of the ground state $|N\rangle$ is 
\[
Q_S=\sum_{n=0}^{N-1}1=N(N-1)/2
\]
which is the largest possible  S charge 
for a given energy $E=N(N-1)/2$. 
The ground state $|N\rangle$ is uniquely characterized 
as  the lowest-energy state satisfying 
\[
Q_S=E 
\]
for a given number $N$ of D3-branes.  
Below we will construct the bilinear operators 
corresponding to all 1/2-BPS operators such that they 
carry zero S-charge. Hence, 
it is possible to consistently 
restrict the physical Hilbert space  in the super-selection 
sector of the fixed S charge $Q_S=N(N-1)/2$ for any given $N$. 
The S-charge is analogous to the U(1) R charge in the 
case of the special operators with $\Delta=J$. 
The spacetime interpretation of the S-charge conservation 
will be discussed later. We will argue that 
it is related to the scale symmetry as the special case of 
conformal symmetry of the ${\cal N}=4$ susy 
Yang-Mills theory, which is apparently lost in 
moving to 1-dimensional matrix models. 

The question is now what is the appropriate algebra for the 
SO(6) vector operators. An obvious guess would be that they 
satisfy the standard bosonic algebra. But this would lead to a  
different normalization $\langle N'|N\rangle =N!$ for the ground state. 
At first sight, such a deficiency could be trivially circumvented by simply 
multiplying the normalization factor $1/\sqrt{N!}$. However, 
it turns out that when we consider general two-point functions of 
general multi-trace operators it actually leads to a wrong relative  
normalization for different partitions of the matrix traces, 
which is not consistent with the fundamental factorization 
property (\ref{factorcomplexrep}) 
and cannot be removed by the change of 
overall normalization. 

Our proposal is that the vector operators satisfy a special kind of 
{\it free} algebra. We introduce the set of operators 
$c^{\dagger}_{i_1i_2\cdots i_n}, c_{i_1i_2\cdots i_n}$ 
with traceless and totally symmetric SO(6) indices, satisfying 
\EQ
c_{i_1i_2\cdots i_n} c_{i_1'i_2'\cdots i_n'}^{\dagger}=\delta_{i_1i_2\cdots i_n, i_1'i_2'\cdots i_n'}, 
\EN
\EQ
\sum_{n=0}^{\infty}{1\over n!}c_{i_1i_2\cdots i_n}^{\dagger} c_{i_1i_2\cdots i_n}=1,  
\EN
where  
$\delta_{i_1i_2\cdots i_n, i_1'i_2'\cdots i_n'} $ symbolically 
designates the identity bi-tensor in the space of traceless symmetric 
tensors of rank $n$. We often denote 
these tensor operators symbolically 
as $c_{(n)}, c_{(n)}^{\dagger}$ suppressing their  indices:
\[
c_{(k)}c_{(\ell)}^{\dagger}=\delta_{(k), (\ell)}, 
\quad \sum_{(k)=0}^{\infty}c_{(k)}c^{\dagger}_{(k)}=1.
\]

This type of algebras is called the Cuntz algebra \cite{cuntz} in the general theory 
of $C^*$ algebras. 
The Cuntz algebra has previously been utilized in field theories 
for constructing `master' fields in the large $N$ limit \cite{haan}. 
More recently, it has also been utilized to study the 
pp-wave limit \cite{bmn}. All these previous applications 
are related to the planar limit. 
In our case, however, 
$N$ is assumed to be arbitrary and, hence, 
the role played by the free nature of this algebra in our construction 
is completely different from those in other applications. Mathematically, 
it is known \cite{abe} that the canonical anticommutative relations 
(CAR)  can be embedded in the Cuntz algebra. This suggests to embed 
the whole algebra of composite fermions 
into a larger Cuntz algebra. But we do not pursue 
such a possiblity in the present work. 
 
We  denote the lowest operator with no SO(6) indices 
({\it i. e.} identity representation) by 
$c_0^{\dagger}$ and $c_0$. 
The vector operators $c_I, c_I^{\dagger}$ discussed previously 
are related to the above by 
\EQ
c_I=w^I_{i_1i_2\cdots i_n}c_{i_1i_2\cdots i_n}\equiv 
w^I_{(n)}c_{(n)}, \quad 
c^{\dagger}_I=w^I_{i_1i_2\cdots i_n}c_{i_1i_2\cdots i_n}^{\dagger}
\equiv w^I_{(n)}c_{(n)}^{\dagger}
\EN
which  lead to 
\EQ
c_Ic^{\dagger}_{I'}=w^I_{i_1i_2\cdots i_n}c_{i_1i_2\cdots i_n}
w^{I'}_{i_1'i_2'\cdots i_n'}c_{i_1'i_2'\cdots i_m'}^{\dagger}=
\delta_{mn}\langle w^{I}w^{I'}\rangle =\delta^{II'}. 
\EN

Although the composite fermion operators themselves 
do not satisfy the CAR, it is easy to check that the 
fermionic nature (\ref{fermionicnature}) of the 
composite operators  when acting upon the 
ground state, and 
hence the correct normalization condition,  are satisfied, 
\[
\langle N'|N \rangle =\delta_{N', N}
(c_0)^N(c_0^{\dagger})^N \langle 0|
b_0 \cdots b_{N-2}b_{N-1}\, b^{\dagger}_{N-1}b_{N-2}^{\dagger}
\cdots b_0^{\dagger}|0\rangle=\delta_{N',N}
\] 
since $(c_0)^N(c_0^{\dagger})^N=(c_0)^{N-1}(c_0^{\dagger})^{N-1}=
\cdots =c_0c_0^{\dagger}=1$. 
Note that, with respect to the Cuntz algebra,  
we extract the internal product as the coefficient of 
the identity. A more legitimate way would be 
to introduce the Fock vacuum 
by $c_I|0\rangle_c=0=\,  _c\langle 0|c_I^{\dagger}$. 
Then, we should write as $_c\langle 0|(c_0)^N(c_0^{\dagger})^N|0\rangle_c=1$. 
In this convention, the second definining equation of the 
Cuntz algebra should be $\sum_Ic^{\dagger}_Ic_I=1-|0\rangle_c 
\, _c\langle 0|$. As long as we consider only the correlation 
function for the groud state, however, our convention, following the 
original form in \cite{cuntz}, 
of not introducing the vacuum state for the 
Cuntz operators is sufficient and simpler.

\section{D-brane fields and their bilinears}
\setcounter{equation}{0}
\subsection{D-brane fields for spherical D3-branes in 1/2-BPS 
sector}

We are now ready to define the field operators of D3-branes in the 
1/2-BPS sector as functions of real coordinates $\phi_i$ and 
a complex coordinate $\alpha$ (complex conjugate 
being $\overline{\alpha}$). 
The fields and their cojugates which annihilate and 
create spherical D3-branes are, respectively,  
\EQA
\Psi_n^{(+)} [\phi, \alpha, \overline{\alpha}]&=&\cases{
\sum_{k=0}^{\infty}\sqrt{{2^{n+k}\over (n+k)!}}\e^{-|\alpha|^2
-|\phi|^2/4}{f(k)\over \sqrt{k!}}\phi_{i_1}\phi_{i_2}\cdots \phi_{i_k}\alpha^{n+k}\, b_{n+k}\, 
c_{i_1i_2\cdots i_k} ,  & $(n\ge 0)$\cr
\sum_{k=-n}^{\infty}\sqrt{{2^{n+k}\over (n+k)!}}\e^{-|\alpha|^2
-|\phi|^2/4}{f(k)\over \sqrt{k!}}\phi_{i_1}\phi_{i_2}\cdots \phi_{i_k}\alpha^{n+k}\, b_{n+k}\, 
c_{i_1i_2\cdots i_k} ,  &  $(n<0)$ \cr}\\
\Psi_n^{(-)} [\phi, \alpha, \overline{\alpha}]&=&\cases{
\sum_{k=0}^{\infty}\sqrt{{2^{n+k}\over  (n+k)!}}\e^{-|\alpha|^2
-|\phi|^2/4}{1\over f(k)\sqrt{k!}}\phi_{i_1}\phi_{i_2}\cdots \phi_{i_k}\overline{\alpha}^{n+k}\, 
b_{n+k}^{\dagger}\, 
c_{i_1i_2\cdots i_k}^{\dagger} ,  &  $(n\ge 0)$ \cr
\sum_{k=0}^{\infty}\sqrt{{2^{n+k}\over  (n+k)!}}\e^{-|\alpha|^2
-|\phi|^2/4}{1\over f(k)\sqrt{k!}}\phi_{i_1}\phi_{i_2}\cdots \phi_{i_k}\overline{\alpha}^{n+k}\, 
b_{n+k}^{\dagger}\, 
c_{i_1i_2\cdots i_k}^{\dagger}  . &  $(n<0)$ \cr}
\EQN
We suppressed the time dependence $b_{n+k}=b_{n+k}(\tau), 
b_{n+k}^{\dagger}=b_{n+k}^{\dagger}(\tau)$. 
These fields are defined such that they have
 definite S-charges, $Q_S=-n$ and $Q_S=n$, 
respectively. In view of the ambiguity, 
$c_{(k)}\rightarrow f(k)c_{(k)}, c_{(k)}^{\dagger}
\rightarrow c_{(k)}^{\dagger}/f(k)$, being  inherent in the 
definition of the Cuntz algebra, we temporalily put an undermined 
$k$-dependent coefficient $f(k)$ which will be fixed later  
such that it gives the correctly normalized correlation 
functions. Unless $f(k)=1$, the fields $\Psi^{(+)}_n$ and 
$\Psi^{(-)}_n$ are {\it not} mutually (hermitian) conjugate to 
each other. 

Though the base space involves 
8 real coordinates and the time $\tau$, 
its {\it effective} spatial dimensions can be  regarded 
as 6,  in the sense that 
the dependence on the six vector coordinates $\phi_i$'s 
is only through traceless polynomials ({\it i.e.} 
spherical harmonics on S$^5$) and the holomorphy 
conditions are satisfied with respect  to the complex coordinate $\alpha$, 
\EQ
(\alpha +{\partial \over \partial \overline{\alpha}})\Psi_n^{(+)}=0=
 (\overline{\alpha}+{\partial \over \partial \alpha})
\Psi_n^{(-)}, \quad 
(\alpha +{\partial \over \partial \overline{\alpha}})\Psi_n^{(-) \dagger}=0=
 (\overline{\alpha}+{\partial \over \partial \alpha})
\Psi_n^{(+) \dagger}. 
\EN  
Roughly, the vector coordinates  should be 
interpreted to be corresponding 
to  $S^5$, while the additional real dimension $x \sim \alpha+
 \overline{\alpha}$ parametrized 
 by the coherent state representation can be interpreted as 
the `radial' direction of AdS$_5$. 
The apparent duplication of the vector coordinates and 
the radial coordinate is related to the composite 
nature of the D-brane creation and annihilation 
operators. In the direct matrix 
language, the extraction of the radial direction has been 
a difficult question. It is interesting that the extended
 fermion picture indicates 
a particular way of extracting the radial direction 
for 1/2-BPS sector. 
It would be desirable to clarify the spacetime 
picture in more geometrical terms.  

   From the viewpoint of ordinary field theories, it seems 
more natural to define fields by using all of independent 
composite creation and annilation operators simultaneously as 
\EQ
\Psi^{(+)}[\phi, \alpha, \overline{\alpha}]\equiv 
\sum_{n=-\infty}^{\infty}\Psi_n^{(+)}[\phi, \alpha, \overline{\alpha}], 
\quad 
\Psi^{(-)}[\phi, \alpha, \overline{\alpha}]
\equiv 
\sum_{n=-\infty}^{\infty}\Psi_n^{(-)}[\phi, \alpha, \overline{\alpha}] .
\EN
Neither of  the fields $\Psi_n^{(+)}, \Psi_n^{(-)}$ with definite S-charge,
nor the $\Psi^{(+)}, \Psi^{(-)}$ without definite S-charge,  
do not satisfy the standard 
canonical (anti-) commutation relations, because of the 
free nature of the Cuntz algebra. Hence, 
they are not local in the sense of the usual framework 
of quantum field theory. 

However, the fields 
$\Psi^{(+)}, \Psi^{(-)}$ without definite S-charge can be 
regarded as a sort of local fields which are mutually 
`quasi-canonical' conjugate to each other. This can be seen by looking at the 
effect of their action on the vaccum:
\EQ
\Psi^{(-)}[\phi, \alpha, \overline{\alpha}]
\Psi^{(+)}[\phi', \alpha', \alpha']|0\rangle 
=|0\rangle \, \delta[\phi, \alpha, \overline{\alpha}; 
\phi', \alpha', \overline{\alpha}']
\EN
where 
\EQ
\delta [\phi, \alpha, \overline{\alpha};\phi', \alpha', \overline{\alpha}']
=\delta[\phi; \phi']\, \delta[\alpha, \overline{\alpha}; 
\alpha',\overline{\alpha}'], 
\EN
\EQ
\delta[\phi; \phi']\equiv 
\sum_{(k)=0}^{\infty}{1\over k!}\phi_{i_1}\phi_{i_2}\cdots \phi_{i_k}
\phi_{i_1}'\phi_{i_2}'\cdots \phi_{i_k}'
\, \e^{-(|\phi|^2 +|\phi'|^2)/4}, 
\EN
\EQ
\delta[\alpha, \overline{\alpha}; 
\alpha',\overline{\alpha}']
\equiv 
\sum_{n=0}^{\infty}
{2^n\over n!}\Big(\overline{\alpha}\alpha'\Big)^n \, \e^{-|\alpha|^2-|\alpha'|^2}, 
\EN
are delta-functions in the spaces of wave functions of the 
vector and radial (holomorphic) 
coordinates, respectively, 
satisfying 
\EQ
\int [d\phi']G[\phi']\, \e^{-|\phi|^2/4}\delta[\phi;\phi']=G[\phi]\, \e^{-|\phi|^2/4}, 
\EN
\EQ
\int d\alpha' d\overline{\alpha}'
F(\alpha')\, \e^{-|\alpha'|^2}
\delta[\alpha, \overline{\alpha}; \alpha', \overline{\alpha}']
= F(\alpha)\, \e^{-|\alpha|^2}, 
\EN
for arbitraty polynomial functions, 
$G[\phi]$ and $F[\alpha]$. 
The integration measure in the space of single-particle wave  functions 
is normalized such that 
\EQ
\int [d^6\phi |d\alpha|^2]\exp(-2\alpha\overline{\alpha}-
|\phi|^2/2)=1,   \quad (|\phi|^2=\sum_{i=1}^6\phi_i^2).
\EN
The origin of the Gaussian measure is of course the free-field nature 
of the matrix model. 
Thus, 
the fields $\Psi^{(+)}, \Psi^{(-)}$ correspond to 
one-particle states 
 which can be strictly localized in the base space. 
However,  this in turn implies that,  
in any allowed quantum state with a {\it definite} value of S-charge,  
D3-branes cannot in general be localized with respect to 
the transverse directions.  
Note also that the fields and their hermitian conjugates 
cannot be local with respect to each other 
in any sense unless $f(k)=1$.

\subsection{Bilinears}
Since we demand that the S-charge is a superselection 
charge, the allowed observables in our Hilbert space 
must have zero S-charge. 
Under this convention, we introduce 
 the integrated bilinear operator which corresponds to a 
single trace operator  
$w^I_{i_1i_2\cdots i_k}\Tr(\phi_{i_1}\phi_{i_2}\cdots \phi_{i_k})$, 
 \[
\int [d^6\phi |d\alpha|^2]\sum_{n=-\infty}^{\infty}
\Psi_n^{(-)}[\phi, \alpha, \overline{\alpha}]
w^I_{i_1i_2\cdots i_k}\phi_{i_1}\phi_{i_2}\cdots \phi_{i_k}
\alpha^k
\Psi_n^{(+)}[\phi, \alpha, \overline{\alpha}]
\]
\EQ
=2^{-k/2}\sum_{n+k_2\ge 0}^{\infty}{\sqrt{(n+k+k_2)!}\over 
\sqrt{(n+k_2)!}}
{f(k_2)\sqrt{k_1!}\over f(k_1)\sqrt{k_2!}} 
 \langle c^{\dagger}_{(k_1)}w^I_{(k)}c_{(k_2)}\rangle
b_{n+k_1}^{\dagger}b_{n+k_2}
\label{bilinear0} ,  \quad (k_1=k+k_2) .
\EN
This is appropriate when the matrix operator corresponds to 
the holomorphic matrix operator in the factorized expression 
using the complex representation (\ref{factorcomplexrep}). 
When the operator corresponds to anti-holomorphic matrix 
operator, this has to be replaced by its hermitian conjugate, 
\[
\int [d^6\phi |d\alpha|^2]\sum_{n=-\infty}^{\infty}
\Psi_n^{(+)}[\phi, \alpha, \overline{\alpha}]^{\dagger}
w^I_{i_1i_2\cdots i_k}\phi_{i_1}\phi_{i_2}\cdots \phi_{i_k}
\overline{\alpha}^k
\Psi_n^{(-)}[\phi, \alpha, \overline{\alpha}]^{\dagger}
\]
\EQ
=2^{-k/2}\sum_{n+k_1\ge 0}^{\infty}{\sqrt{(n+k+k_1)!}\over 
\sqrt{(n+k_1)!}}
{f(k_1)\sqrt{k_2!}\over f(k_2)\sqrt{k_1!}} 
 \langle c^{\dagger}_{(k_1)}w^I_{(k)}c_{(k_2)}\rangle
b_{n+k_1}^{\dagger}b_{n+k_2}
\label{bilinear0bar} ,  \quad (k_2=k+k_1) .
\EN
It will turn out that for the correct normalization 
of correlators the choice 
\EQ
f(k)=\sqrt{k!}
\label{f}
\EN
is most appropriate, such that the fermion part 
of the bilinear operators take the same form 
as those obtained from the comlex 1-matrix model and 
hence the factorization property of the correlators are 
faithfully reallized.  

These purely holomorphic and anti-holomorphic operators 
are sufficient to compute all correlators of the extremal type 
which include the two-point functions. 
In the non-extremal case, the identification of (\ref{bilinear0}) 
or its conjugate (\ref{bilinear0bar}) 
with the single-trace matrix operator $w^I_{i_1i_2\cdots i_k}\Tr(\phi_{i_1}\phi_{i_2}\cdots \phi_{i_k})$ is not sufficient 
for properly taking into account the normal ordering prescription, 
since then the bilinear operators of mixed type become necessary. 
In this first work, we restrict ourselves only to the extremal case. 
 
As a special case of general bilinear operators that correspond to 
the case $k=0$, 
the Hamiltonian is expressed as 
\[
H=\int [d^6\phi |d\alpha|^2]\sum_{n=-\infty}^{\infty}
\Psi_n^{(-)}[\phi, \alpha, \overline{\alpha}]
(\alpha{\partial \over \partial \alpha} +\alpha \overline{\alpha})
\Psi_n^{(+)}[\phi, \alpha, \overline{\alpha}]
\]
\EQ
=\Big(\sum_{(k)=0}^{\infty}c_{(k)}^{\dagger}c_{(k)}\Big)
\Big(\sum_{n=0}^{\infty}nb_n^{\dagger}b_n\Big)
=\sum_{n=0}^{\infty}nb_n^{\dagger}b_n, 
\EN
which is automatically 
hermitian for arbitrary $f(k)$. 
We note that actually the Hamiltonian can also be expressed 
in the standard local form using the fields without 
definite S-charge as 
\EQ
H=\int [d^6\phi |d\alpha|^2]
\Psi^{(-)}[\phi, \alpha, \overline{\alpha}]
(\alpha{\partial \over \partial \alpha} +\alpha \overline{\alpha})
\Psi^{(+)}[\phi, \alpha, \overline{\alpha}]. 
\EN
Since this behaves as the identity with respect to the 
Cuntz  algebra, the Heisenberg equation 
of motion is consistent with the time dependence that 
has been assumed in the foregoing discussions.  
Similarly, the number operator is 
\[
N=\int [d^6\phi |d\alpha|^2]\sum_{n=-\infty}^{\infty}
\Psi_n^{(-)}[\phi, \alpha, \overline{\alpha}]
\Psi_n^{(+)}[\phi, \alpha, \overline{\alpha}]
\]
\EQ
=\Big(\sum_{(k)=0}^{\infty}c_{(k)}^{\dagger}c_{(k)}\Big)
\Big(\sum_{n=0}^{\infty}b_n^{\dagger}b_n\Big)
=\sum_{n=0}^{\infty}b_n^{\dagger}b_n. 
\EN
This can also be expressed in the local form 
in terms of $\Psi^{(+)},\Psi^{(-)}$. 
 
Finally, for purely `kinematical' operators as the 
bilinears of the Cuntz operators alone are represented as 
{\it ratios} of the ordinary bilinears. For instance, 
we can define
\EQ
R\equiv \sum_{k=0}^{\infty}k\, c^{\dagger}_{(k)}c_{(k)}
=N^{-1}
\int [d^6\phi |d\alpha|^2]
\Psi^{(-)}[\phi, \alpha, \overline{\alpha}]
\Big(
\phi_i{\partial \over \partial \phi_i}+{1\over 2}|\phi_i|^2\Big)
\Psi^{(+)}[\phi, \alpha, \overline{\alpha}], 
\EN
where the expression $N^{-1}$ is meant that it acts 
upon arbitrary 1/2-BPS states except for the Fock vacuum.  
This operator counts the number of transverse scalar fields $\phi_i$ 
and satisfies
\EQ
R\Psi^{(-)}=\Big(\phi_i{\partial \over \partial \phi_i}
+{1\over 2}|\phi_i|^2\Big)\Psi^{(-)}, 
\quad 
\Psi^{(+)}R=\Big(\phi_i{\partial \over \partial \phi_i}
+{1\over 2}|\phi_i|^2\Big)\Psi^{(+)}. 
\EN
The S-charge operator is then given by
\EQ
Q_S=H-R, 
\EN
whose commutator with a bilinear counts its S-charge. 
It is therefore commutative with the above (\ref{bilinear0}) and 
(\ref{bilinear0bar}). Note that to compute the commutator of 
R and general bilinears, the above two relations are sufficient:
$
[R, c_{k_1}^{\dagger}c_{k_2}]=(k_1-k_2)c_{k_1}^{\dagger}c_{k_2}. 
$
We can similarly construct SO(6) generators as, 
\EQ
J_{ij}\equiv N^{-1}
\int [d^6\phi |d\alpha|^2]
\Psi^{(-)}[\phi, \alpha, \overline{\alpha}]
\Big(
\phi_i{\partial \over \partial \phi_j}-\phi_j{\partial \over \partial \phi_i}\Big)
\Psi^{(+)}[\phi, \alpha, \overline{\alpha}], 
\EN
satisfying 
\EQ
J_{ij}\Psi^{(-)}=-\Big(
\phi_i{\partial \over \partial \phi_j}-\phi_j{\partial \over \partial \phi_i}\Big)
\Psi^{(-)}, \quad 
\Psi^{(+)}J_{ij}=\Big(
\phi_i{\partial \over \partial \phi_j}-\phi_j{\partial \over \partial \phi_i}\Big)
\Psi^{(+)}. 
\EN
The SO(6) transformations of bilinears are given by taking commutator  with $J_{ij}$.

\subsection{Two-point functions}

Let us now check that these S-charge invariant bilinear 
operators give correct correlation functions. 
In what follows, we always assume that the operators are 
time-ordered with respect to the Euclidean time $\tau$.  
For  two-point functions of  single-trace operators, 
\[
\langle w^{I_1}_{i_1i_2\cdots i_k}\Tr(\phi_{i_1}\phi_{i_2}\cdots \phi_{i_k})
(\tau_1)\, 
w^{I_2}_{j_1i_2\cdots j_k}\Tr(\phi_{j_1}\phi_{j_2}\cdots \phi_{j_k})
(\tau_2)\rangle, 
\]
only the terms with $k_2=0, k_1=k$ or $k_1=0, k_2=k$ of 
the above bilinear operators 
contribute, since the action of the vector creation or annihilation 
operators on the ground state is nonvanishing only 
for the trivial representation. 
Here we recover the time dependence, 
$\exp\Big(-k(\tau_1-\tau_2)\Big)$. 
Thus the  extended 
fermion representation of the two-point function  with the 
choice (\ref{f}) is equal to 
\[
\langle N|\langle c^{\dagger}_0w^{I_1}_{(k)}c_{(k)}\rangle 
\langle c^{\dagger}_{(k)}w^{I_2}_{(k)}c_0\rangle 2^{-k}
\sum_n\sqrt{{(n+k)!\over n!}}b^{\dagger}_nb_{n+k}
\sum_m \sqrt{{(m+k)!\over m!}}b^{\dagger}_{m+k}b_m |N\rangle
\exp\Big(-k(\tau_1-\tau_2)\Big)
\]
\EQ
=2^{-k}\langle w^{I_1}_{(k)}w^{I_2}_{(k)}\rangle 
 \langle N| \sum_n\sqrt{{(n+k)!\over n!}}b^{\dagger}_nb_{n+k}
\sum_m \sqrt{{(m+k)!\over m!}}b^{\dagger}_{m+k}b_m |N\rangle
\exp\Big(-k(\tau_1-\tau_2)\Big)
\EN
This is precisely the required form satisfying the factorization 
property (\ref{factorcomplexrep}). 

For the multi-trace operator ${\cal O}^I_{(k_1, k_2, \ldots, k_n)}$, 
the corresponding operator acting upon the 
ground state is, when it corresponds to the holomorphic 
matrix operator, 
\EQ
w^I_{(k_1+k_2+\cdots +k_n)}
B_{(k_1)}B_{(k_2)}\cdots B_{(k_n)}, \quad (k_1+k_2+\cdots +k_n=k)
\EN
where $B_{(k_a)}$'s are the bilinear operators with completely 
symmetrized  tensor indices of rank $k_i$, 
\EQ
B_{(k_a)}=\int [d^6\phi |d\alpha|^2]\sum_{n=-\infty}^{\infty}
\Psi_n^{(-)}[\phi, \alpha, \overline{\alpha}]
\phi_{(i_1}\phi_{i_2}\cdots \phi_{i_{k_a})}
\alpha^k
\Psi_n^{(+)}[\phi, \alpha, \overline{\alpha}]. 
\EN
Remember that in the above expressions the tensor indices 
are symbolically represened by the lower subscript such as 
$(k)$. Since the indices are  contracted with $w^I$, 
$B_{(k_a)}$ takes the same form as 
(\ref{bilinear0}) by replacing the 
$w$-tensor in the latter by the part of tensor 
indices $(k_i)$.   If this string of the bilinears 
acts upon the ground state, 
each factor 
 shifts the rank of the vector operators successively as 
$0\rightarrow k_n \rightarrow k_n+k_{n-1} \rightarrow 
\cdots \rightarrow k_n+k_{n-1}+\cdots +k_1=k$ (from right to left). 
Other types of shiftings do not contribute to the 
extremal correlators. 
In this way,  we arrive at  the following product of fermion bilinears 
that acts  upon the ground state, 
\[
2^{-k/2}\langle c_{(k)}^{\dagger}w^I_{(k)}c_0\rangle 
\sum_{\ell_1=0}^{\infty}\sqrt{{(\ell_1+k_1)!\over 
\ell_1!}}b^{\dagger}_{\ell_1+k_1}
b_{\ell_1} 
\cdots \times
\]
\EQ
\cdots \sum_{\ell_{n-1}=0}^{\infty}
\sqrt{{(\ell_{n-1}+k_{n-1})!\over \ell_{n-1}!}}
b^{\dagger}_{\ell_{n-1}+k_{n-1}}b_{\ell_{n-1}}
\sum_{\ell_n=0}^{\infty}
\sqrt{{(\ell_n+k_n)!\over \ell_n!}}b^{\dagger}_{\ell_n+k_n}
b_{\ell_n}|N\rangle\,  \e^{k\tau_2}.
\label{geneprodofbilinear}
\EN
Although the field operators do not satisfy any simple 
commutation relations among themselves, 
their bilinears $B_{(k_i)}$
 give a commutative algebra after acting upon the 
ground state.  

Similarly, we obtain the conjugate operator acting to the left upon 
$\langle N|$ with time dependent factor 
$\e^{-k\tau_1}$. This result is valid for arbitrary partition 
of the traceless symmetric tensor indices of the $w$-tensors. 
They give the required factorized expressions  of 
the form (\ref{factorcomplexrep}) 
with correct normalization for arbitrary two-point functions 
of multi-trace operators. 

To obtain the correctly normalized expressions, 
the Cuntz algebra $c_{(k)}c^{\dagger}_{(\ell)}=\delta_{(k), (\ell)}$ 
is crucial. If we assumed the usual bosonic 
algebra,  the matrix operators 
with different partitions of the vector indices would have given  
differently normalized expressions for the product of fermion bilinears, 
depending on the manner of partitions.  
That would not be compatible with the factorization property 
(\ref{factorcomplexrep}).  
In other words, they would reduce to wrongly normalized 
products of fermion bilinears even when we consider the 
special 1/2-BPS operators with a single charge 
satisfying the condition $\Delta=J$. 
The origin of such discrepancy  is the same as the one 
which leads to the wrong normalization of ground state 
$|N\rangle$, as we  have mentioned previously. 

\subsection{Higher-point extremal correlators}

The case of higher-point extremal correlators is 
essentially understood from the structure of two-point 
functions. The term `extremal' means that 
the conformal dimensions of the operators satisfy 
$\Delta_1=\Delta_2+\Delta_3 +\cdots +\Delta_n$. 
The free-field contractions obviously give the 
following structure
\EQ
\langle {\cal O}_1(x_1){\cal O}_2(x_2)\cdots 
{\cal O}_n(x_2)\rangle
=\langle w_1w_2\cdots w_n\rangle \, 
G(\{(1),(2), \ldots, (n)\};N)\prod_{A=2}^n 
{1\over |x_1-x_A|^{2\Delta_A}}.
\EN
It has been cojectured, with ample evidence just 
as for 2- and 3-point correlators,  that this form is 
not renormalized from free-field result. 

As in the case of two-point correlators, 
the factorization into 1-single matrix models 
owes to the uniqueness of the SO(6) 
invariant. The function $G(\{(1),(2), \ldots, (n)\};N)$ 
is again given by the one-matrix hermitian model, 
replacing the SO(6) matrices $\phi_i$ by 
the single 1-dimensional hermitian matrix field 
$M(\tau)$, with normal ordering condition being 
understood in each operator.
\EQ
\langle {\rm:}{\cal O}_1(\tau_1){\rm :}{\rm :}{\cal O}_2
(\tau_2){\rm :}\cdots {\rm :}
{\cal O}_n(\tau_2){\rm :}\rangle_M
=G(\{(1),(2), \ldots, (n)\};N)
\, \prod_{A=2}^n\, \e^{-\Delta_A(\tau_1-\tau_A)}
\EN
As before, the correspondence of the spacetime factor 
is given by $\e^{\tau_1-\tau_A}\leftrightarrow |x_1-x_A|^2$. 

Here, we have assumed, without loss of  generality, 
 that the operator ${\cal O}_1(\tau_1) 
\leftrightarrow {\cal O}_1(x_1)$ is located at the largest 
time ($\tau_1>\tau_2>
\cdots >\tau_n$) by choosing the origin of the base spacetime 
coordinates of the super Yang-Mills theory appropriately.  
Then, the normal-ordering prescription is handled by 
converting to the complex 1-matrix model in the same way as for 
the two-point case: 
 all of the operators except for 
the first one  ${\cal O}_1(\tau_1)$ is consisting of the 
holomorphic matrix $Z$, while the first one is 
chosen to be the anti-holomorphic operator 
consisting of the conjugate matrix $Z^{\dagger}$.  
Other time-orderings are of course possible, but 
handling of the normal ordering prescription would 
become more cumbersome. 
The situation is now almost the same as in the 
two-point case. Only difference is the 
time dependence, resulting in 
the correct factor $\prod_{A=2}^n\, 
\e^{-\Delta_A(\tau_1-\tau_A)}$. 
It is now clear that the extremal correlators 
can be reproduced by our generalized fermion 
picture in the same way as in the two-point functions. 

\vspace{0.2cm}
\noindent
{\it A short comment on 3-point non-extremal case}

There are good reasons to believe
 (see \cite{nonrenormal}) that the nonrenormalization property 
is valid for 3-point functions including non-extremal cases. 
Therefore it is natural to try to extend our formalism to 
3-point functions. In fact, it is not so difficult to do so 
for some simple cases. One complication is that 
 the normal ordering prescription requires more 
complicated procedures. 
For non-extremal 3-point functions, we have to 
introduce `mixed' bilinear operators, being located at the middle time, 
 which have dependence on both $\alpha$ and $\overline{\alpha}$, 
when the bilinear operators are not directly 
acting upon the ground state. For such operators, 
we have to explicitly subtract the contributions 
from contractions inside the matrix operators. 
In spite of this, 
it is possible to show that the OPE coefficients can be correctly 
reproduced by extending the present approach. 

Another problem which is more fundamental is that our 
`spherical' approximation in which 
the space-time dependence of correlators is obtained by  
a simple substitution $\e^{|\tau_1-\tau_2|}\rightarrow 
|x_1-x_2|^2$ is no more sufficient to reproduce 
the spacetime dependence of non-extremal 
3-point functions. To give a satisfactory 
treatment, it is therefore necessary to extend our 
formalism such that non-spherical modes, with respect to the base space of D3-branes, 
are taken into account appropriately. 
We postpone such elaborations to later works.

\subsection{The meaning of the S-charge symmetry}
Let us here discuss the spacetime meaning of the 
S-charge superselection rule. The product of bilinear operators 
in the expression (\ref{geneprodofbilinear}) 
is, by construction, invariant under the S-charge transformation 
\[
c^{\dagger}_{(k)}\rightarrow \e^{-ik\theta}c^{\dagger}_{(k)}, \quad 
c_{(k)}\rightarrow e^{ik\theta}c_{(k)}, \quad 
b_n^{\dagger}\rightarrow \e^{in\theta}b_n^{\dagger}, \quad 
b_n\rightarrow \e^{-in\theta}b_n
\]
which is generated by the operator $Q_S=H-R$. 
If one `Wick'-rotates the angle,  $\theta\rightarrow i\sigma$ ($\sigma=$ 
real), this is equivalent to the  scaling that  
transforms simultaneously the transverse vectors $\phi_i$'s and the time variable as 
\EQ
\phi_i \rightarrow \lambda \phi_i, \quad 
\e^{\tau_2} \rightarrow \lambda^{-1}e^{\tau_2} \quad 
(\lambda \equiv e^{\sigma}), 
\EN
{\it when} one expresses the transformation 
in terms of the original matrix field variables. 
For the conjugate operator, the transformation is expressed as
\EQ
\phi_i \rightarrow \lambda \phi_i, \quad 
\e^{-\tau_1} \rightarrow \lambda^{-1}e^{-\tau_1} \quad 
(\lambda \equiv e^{-\sigma}).
\EN
Thus, the states constructed by acting the 
bilinears, and hence the correlation functions,  must be invariant under 
these scalings. This is precisely equivalent to the scaling
 symmetry 
of two-point (and general extremal) correlation functions, since 
in terms of the original base spacetime 
coordinates $x_i$ $(i=1, 2)$ of the super Yang-Mills theory, 
\[
\e^{\tau_1-\tau_2}=|x_1-x_2|^2 \rightarrow 
\lambda^{-2}\e^{\tau_1-\tau_2}=
\lambda^{-2}|x_1-x_2|^2. 
\] 
As argued in \cite{jevyo}, behind the (generalized) 
scaling symmetries, 
a dual uncertainty relation of spacetime \cite{stu} characterizes the 
dynamics of D-branes: 
$\Delta T \Delta X \gtsim \alpha'$ with $\Delta T$ being the 
typical microscopic scale in the longitudinal direction $\sim x^{\mu}$ 
along branes, 
and $\Delta X$ being the scale in the transverse directions, 
$X \sim \phi_i$. The scale transformation
$\Delta T\rightarrow \lambda^{-1} \Delta T, 
\Delta X \rightarrow \lambda \Delta X$ changes 
these two characteristic scales, but the changes are  consistent with 
the existence of an uncertainty relation of the above type. 
As emphasized in the first part of the  present section, 
the S-charge superselection rule leads to a 
certain nonlocality with respect to six effective 
spatial dimensions. Since the nonlocality is not directly 
characterized by the form of wave functions, which 
are ordinary functions of the coordinates as in the 
usual local field theory, nor by the form of the Hamiltonian 
which is a local bilinear function,  the above characterization by the 
spacetime uncertainty relation seems quite appropriate. 
Of course, the precise nature must further be clarified.

\section{Multi-charge geometries: Superstar 
and its entropy}
\setcounter{equation}{0}
\subsection{Dexclusion principle and holographay}

From the viewpoint of AdS/CFT correspondence, our 
discussions so far are from the side of the boundary CFT 
theory, to the extent that
 we have been trying to construct a 
second-quantized theory of super Yang-Mills theory 
in the 1/2-BPS sector. One of the characteristics 
of our construction is that the quantum statistics 
of D-brane creation and annihilation operators 
must be assumed to satisfy an extended version
 of Pauli principle. 
Thus, a natural question is now what is the interpretation of the Dexclusion principle from the standpoint
 of bulk supergravity theory. 
 
 \begin{center}
\begin{figure}
\begin{picture}(200,120)
\put(100,0){
\includegraphics[width=230pt]{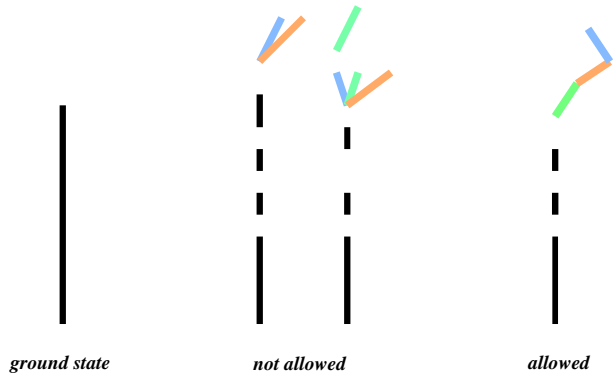}}
\end{picture}
\caption{The Dexclusion principle from  the bulk viewpoint: 
The  droplets are in general deformed when particles are 
excited with non-singlet SO(6) wave functions, but must have 
always the fixed constant density with a single sheet. 
They cannot  bifurcate into two or more sheets. 
The lines show sections of droplet seen when cut by an appropriate 
plane intersecting with the droplets: black=singlet, color (gray)=nonsinglet. 
Compare with Fig. 1.
}
\label{fig2}
\end{figure}
\end{center}

\vspace{-1.3cm}
If we choose different directions in breaking SO(6) 
correponding to a partcular angular momentum $J$, 
the two-dimensional plane of the droplet in the LLM 
classification is deformed 
by various SO(6) rotations. Generic solutions 
generated by those SO(6) rotations have nonzero charges 
with respect to three different SO(2) directions. 
 However,  the basic property of 
 droplet that it is incompressible 
is preserved by such deformations, 
so that the density of the droplets must always be constant 
with the same density as in the case of single charge 
solutions. 
The sheets extending into 
different directions in the bulk would correspond to 
different SO(6) states of (giant) graviton configurations. 
This implies that the two-dimesional sheet of the droplet 
would always be consisting of one sheet. In other words, 
the sheet of the droplet would not bifurcate into 
two or more sheets. 

That this is a natural bulk 
correspondent of the Dexclusion property 
can be easily understood 
by imagining a situtation 
where the DEX is not satisfied 
and hence we have 
excitations of two different SO(6) 
states simultaneously  
with the same energy. Then, the corresponding situation 
on the bulk side would look like this: 
the sheet defining the 
ground state is deformed, after this excitation, 
 into a configuration which 
bifurcates somewhere near the boundary of the 
ground-state droplet into two independent sheets
extending in different directions.
See Fig. \ref{fig2} in comparison with Fig. \ref{fig1}.  
In fact, it is very difficult to imagine 
that the set of smooth classical 
solutions exhibit such a singular behavior as bifurcation. 
Thus, from the viewpoint of the holographic correspondence 
between bulk and boundary theories, the Dexclusion 
principle seems to be a natural extension of 
the usual Pauli principle 
in formulating D-brane field theory. 

It would be very interesting to formulate this qualitative 
picture in a more concrete form on the bulk side 
by extending the 
analysis of LLM \cite{LLM} 
to more general configurations with smaller 
isometries, so that we could 
 treat `covariantly' all of 1/2-BPS SO(6) states  
 with multiple SO(2) charges. 
It would require us to extend their ansatz in such a 
way that, with respect to the 
isometry SO(4)$\times$SO(4)$\times$ {\bf R}, 
only the first SO(4) isometry corresponding to 
the spherical approximation of D3-branes is retained 
while the remaining factors are replaced by some 
appropriate form which allows inclusion of all 
possible SO(6) states corresponding to the 
representation $(0, k, 0)$.

\subsection{Superstar entropy}
\begin{center}
\begin{figure}
\begin{picture}(200,160)
\put(100,0){
\includegraphics[width=260pt]{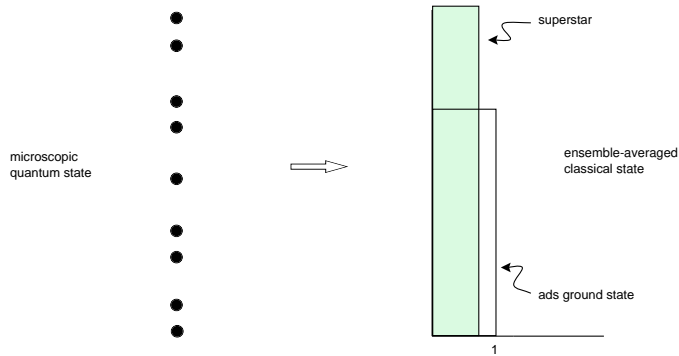}}
\end{picture}
\caption{The microstates (left) of a superstar and the corresponding 
averaged configuration (right) represented by a droplet with 
a lower density. 
}
\label{fig3}
\end{figure}
\end{center}

At present, only solutions whose form is explicitly 
known with a more general SO(6) 
configurations are the so-called 
 `superstar' solutions \cite{mytaf}. In the extremal limit, 
they in general  have (naked) singularities at their center. 
It has been shown that they have some characteristics which 
can be interpreted in terms of the condensation of giant gravitons. 
It is interesting, as is already noted by 
LLM,  that in the single charge case satisfying the 
condition $\Delta=J$, a superstar corresponds to a 
choice of droplet with a density  lower 
 than the smooth solutions. 
The lower density of a superstar droplet could be 
intepreted as a statistical average 
of loosely packed occupied states\footnote{ 
This is reminiscent of the situation in fractional quantum Hall effect,  
to which an analogy was already alluded in section 3 
from a different  aspect. }of fermions. 
There exist a large number of nonsingular 
solutions  with the same charge and the same mass 
as a singular superstar solution, but with different 
microscopic distributions of occupied states. 
Although the existence of both 
nonsingular and singular solutions with the same 
charges is mysterious and hence
 precise relation between the nonsingular and 
singular solutions must be clarified, it seems natural 
to view a superstar as an approximation to the ensemble of a large 
number of droplet configurations  corresponding to 
nonsingular LLM solutions, as in Fig. \ref{fig3}. 
The apparent singularity of the superstar solution  
would be smoothed out by stringy quantum effects, 
as observed in other situations \cite{smallblack}. 
The entropy of a general superstar would then be obtained 
by counting microstates corresponding to the 
nonsingular LLM solutions \cite{count, balasu3}. 
Let us make a modest 
consistency check on the entropy of superstar from this viewpoint.
This viewpoint is in accord with the so-called `Fuzzball' conjecture 
\cite{mathur}.  Similar ideas as above for superstars 
 have recently been discussed for the case 
of D1-D5 system in the literature \cite{balas2}.   
Even Schwarzschild-type  black holes without 
any supersymmetry might be understood in a similar  way \cite{balasu3}.   

\vspace{-1.4cm}
The generic superstar solutions that have smaller 
symmetries than $SO(4)\times SO(4)\times {\bf R}$ 
have three independent $U(1)$ charges ($J_1, J_2, J_3$) 
corresponding to $U(1)^{3}$ embedded in the 
R-symmetry group $SU(4)\sim SO(6)$. The energy (mass) equals 
to the sum 
\[
\Delta=|J_1|+|J_2|+|J_3|,
\]
 which might 
naively be interpreted to be an indication that the 
microscopic degrees of freedom could be independently 
distributed over the 3 directions of U(1)$^3$.  
According to our extended fermion field 
theory, we can make a simple prediction for the 
entropy of generic superstar. 
The generic quantum states take the factorized form 
(\ref{geneprodofbilinear}) for a given SO(6) 
wave function. Therefore,  provided that 
the  entropy of  the single charge superstar 
is given as 
\[
S_{single}=f(|J|), 
\]
the entropy of three charge superstar takes the form
\EQ
S_{multi}=f(|J_{1}|+|J_{2}|+|J_{3}|), 
\EN
using the same function $f(J)$. 
This itself is a consequence of SO(6) symmetry 
since the entropy must be invariant under SO(6) 
and $k\equiv |J_{1}|+|J_{2}|+|J_{3}|$ is the Dynkin label 
specifying the SO(6) representation. However, the Dexclusion pricniple (DEP) shows that the particles with different 
charges cannot be excited independently, and therefore 
the number of microstates must be smaller than 
the case of independent excitations.  This means that 
the inequality
\EQ
f(|J_{1}|+|J_{2}|+|J_{3}|)<f(|J_{1}|)+f(|J_{2}|)+f(|J_{3}|)
\label{entropy}
\EN
must be satisfied when at least two of the angular momenta are not zero. If we have assumed naively that the mircoscopic degrees of 
freedom were distributed independently for different 
directions, (\ref{entropy}) must be replaced by equality, and hence 
we  would be lead to a {\it wrong} prediction that $f(J)\propto J$.  

Consider the partition function of fermion spectrum ($u, v <1$) 
\EQ
Z(u, v)=\Tr\Big(u^Nv^H\Big)=\sum_{n=0}^{\infty}
d_n(v) u^n=\sum_{n,m=0}d_{n,m}v^{m+{n(n-1)\over 2}}u^n .
\EN
The entropy of the superstar with the total U(1)$^3$ charge $\Delta$ 
and total RR-charge $N$ 
is given as 
\EQ
S=\ln d_{N, \Delta}. 
\EN
Using the product formula 
\[
\prod_{n=0}^K
(1+uv^n)=
1+ \sum_{r=1}^{K+1}
{(1-v^{K+1})(1-v^{K})\cdots (1-v^{K+1-r+1})
\over (1-v)(1-v^2)\cdots (1-v^r)}
\, v^{r(r-1)/2}u^r, 
\]
we find, after taking the limit $K\rightarrow \infty$, 
\EQ
d_n(v)=v^{n(n-1)/2}\prod_{r=1}^n{1\over 1-v^r}, 
\EN
which can also be directly derived from the matrix representation 
in the single-charge case. 
This enable us to 
estimate the entropy 
in the limit of large $N$ and $\Delta$ satisfying 
$\Delta /N^2\ll 1, N\gg 1, \Delta\gg 1$, as \cite{count}
\EQ
f(\Delta) \sim \Big({2\pi^2\over 3}\Big)^{1/2}\sqrt{\Delta}, 
\EN
which indeed satisfies the above inequality. In the opposite regime 
$\Delta/N^2\gtsim 1 \, (N\gg 1)$, the particle distribution 
becomes very dilute and 
hence the above  inequality will almost be saturated with some 
mild corrections. 
It would be desirable to extend the present
 analysis to arbitrary finite $N$ and $\Delta$. 
 For a more detailed study of superstar entropy for 
 the single charge case, see \cite{count, balasu3}.

\section{Discussions}
\setcounter{equation}{0}
To summarize, with the motivation of constructing 
a quantum field theory for D-branes in mind, 
we have first critically 
reviewed the known fermion representation of 1/2-BPS operators. 
We have argued the importance of treating all directions of 
scalar  fields on an equal footing.  
It was shown that this can indeed be achieved 
by the help of non-renormalization property for the extremal 
correlators of 1/2-BPS operators.  
We have then proposed a field theoretic representation  
of the multi-body system of spherical D3-branes in 1/2-BPS sector. 
There are two characteristic properties of this extended 
fermion field theory. One is the necessity of extending Pauli 
principle to a much more stronger version, coined as 
`Dexclusion principle'. 
This seems to be an inevitable consequence from the 
factorizable structure of the extremal correlators. This also seems to be  
consistent with holography, since it has a natural qualitative interpretation 
on the supergravity side. 
Another is a rather peculiar nonlocality, intimately related with 
a superselection rule which can be interpreted as a 
disguise of the scale symmetry. As such, the nonlocality 
conforms to the spacetime uncertainty principle, 
though in the present approximation stringy degrees of 
freedom do not play roles in any manifest manner. 

\begin{center}
\begin{figure}
\begin{picture}(180,140)
\put(100,20){
\includegraphics[width=240pt]{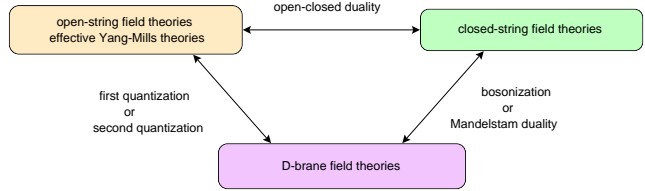}}
\end{picture}
\caption{This illustates the conceptual relation of 
a possible framework of D-brane field theory with 
the usual open and closed string-field theories. }
\label{fig4}
\end{figure}
\end{center}

\vspace{-1cm}
From the viewpoint of D-brane field theory, 
what we have done 
 remains  yet at a very modest level. 
It is not clear at this stage 
whether our construction could really be a 
starting point towards a more general framework, or 
merely shows a specialty restricted to the 1/2-BPS 
sector. If this approach is successful, it 
would ultimately provide third possible formulation of 
string theory which could connect open and closed string field theories from a new perspective, providing a new 
explanation of open-closed string duality: 
On one hand, it would be a second-quantization 
of open string-field theory, while on the other hand 
it would be related to closed-string field theory 
by the Maldelstam-type duality. See Fig. \ref{fig4}. 
It is a great challenge to develop 
further our formalism to a level where 
we could envisage such possibilities.

There are many possible directions related to the 
results of the present work. Let us conclude by listing some of them 
that have not been mentioned in previous sections. 

\vspace{0.2cm}
\noindent (1)\, \,   First of all, our formalism itself looks somewhat artificial yet, for lack of firm principles 
from which we could hopefully deduce a much tighter structure 
of the theory uniquely. 
The complex matrix model has been argued not to be 
essential for the  emergence of the generalized fermion 
picture, but still has been used as a technical device for dealing with 
the normal ordering prescription. The method 
works well for extremal correlators. But it becomes 
very complicated,  once we try to treat non-extremal 
correlators. There may be better formulation(s) using more 
appropriate languages for this purpose. 

\vspace{0.2cm}
\noindent (2)\, \, Extension to non-sherical branes and 
to a satisfactory treatment of general non-extremal correlators is also 
an important technical problem which must be resolved 
to  make real progress along the present approach. 
Related to this is the problem of supersymmetric completion 
of the theory. One of the crucial question would be 
what is the appropriate generalization of the S-charge symmetry 
for non-spherical degrees of freedom. Since the S-charge 
transformation can be interpreted as a sort of 
`unitary trick' of the scale transformation, a certain 
unitary version of SO(4,2) group and its supersymmetric 
extension might be important. 

\vspace{0.2cm}
\noindent (3)\, \,  Extension to non-BPS operators as well as to 
other BPS operators \cite{ber2} with smaller supersymmetry:
The treatment of full stringy degrees of freedom  would perhaps require  an extension of  the base space into some kind of 
 noncommutative space. It is also an interesting question how 
 the DEX could be extended (or reinterpreted) 
 for non-BPS states which are 
 characterized by anomalous conformal dimensions. 
We could begin from some appropriate 
approximations, such as, for instance, the PP-wave limit \cite{bmn}. 
The well known relation with spin chains \cite{spinchain} 
would also be useful. 
The extension to non-BPS modes will be a prerequisite for the 
inclusion of anti D-branes in the formalism. 

This also raises the 
problem of possible nonlinear realization of the maximal ${\cal N}=2$ 
spacetime supersymmetry in 10 dimensions, as 
we have emphasized in \cite{yosu} in a more general 
context. In particular, it is pointed out 
there that the realization of full 10D ${\cal N}=2$ supersymmetry in the presence of both D- and anti-D-branes 
is closely woven  with the  open-closed string duality and 
also with the $s$-$t$ channel duality. 
As stressed above, D-brane field theory is expected to provide 
a new  bridge between open and closed strings. It would be 
quite worthwhile to revisit the problem from the viewpoint of 
D-brane field theory. 

\vspace{0.2cm}
\noindent (4)\, \,  In  connection to the PP-wave limit, we would like to 
add a further remark. In the LLM classification, the 
 circle at the boundary of the ground state 
droplet, sitting just at the center ($\rho=0$) of the AdS$_5$ spacetime,  
coincides with the geodesic which is usually used for studying the 
PP-wave limit.  For the purpose of establishing the holographic 
correspondence of correlators, it is useful \cite{dsy} to euclideanize the 
geodesics by performing a Wick rotation for the global time and 
simultaneously for the angle parametrizing a large circle of S$^5$. 
The euclideanized geodesic starts from a 
point on the AdS boundary and go back to another point of 
the AdS boundary. This makes possible to directly take the 
BMN limit of two- and three- point correlation functions 
through the GKP-Witten relation.\footnote{
We warn the reader that this Wick rotation cannot be 
done as an analytic continuation from the Minkowski 
formulation, rather should be taken to be a 
requisite for the possibility of directly taking the BMN limit if we wish to preserve  
the GKP-Witten relation.  
} After the  above Wick rotation, 
the droplet becomes non-compact, with its boundary being a 
hyperbola which is the euclidenized geodesic 
itself and asymptotically reaches the (Euclidean) AdS boundary. 
The situation is now quite akin to the $c=1$ matrix model. 
Indeed, the Wick rotation of the angle  parameter 
effectively changes the sign of harmonic oscillator potential. 
The scattering amplitude of ripples traveling along the hyperbola should be related to the correlation functions by a definite 
holographic relation as we have given in \cite{dobayo}. 
It would be very interesting to connect our D-brane field theory 
to the `holographic' string-field theory constructed there 
by bosonization on the basis of this viewpoint. Of course, 
we have to extend the usual method \cite{boson}\cite{maoz} 
of non-relativistic 
bosonization to include transverse coordinates. 

\vspace{0.2cm}
\noindent (5)\, \, An important question related to this last question
 is the problem of so-called leg factors, which are usually necessary 
in making connection between bulk closed strings 
and matrix models. The recent work \cite{donos}  
seems to be very suggestive in this regard. They have 
studied a collective-field treatment of matrix operators 
involving both $Z$ and $Z^{\dagger}$ in the free-field 
approximation. From our point of view including all 
SO(6) directions on an equal footing, these operators 
should rather be related to the 1/2-BPS operators of the following type
\EQ
\Tr\Big(
4Z^{J+1}Z^{\dagger} -\sum_{r=1}^J\phi_iZ^r\phi_iZ^{J-r}
\Big), 
\label{zzbarsinglet}
\EN
which has conformal dimension $\Delta=J+2$ and is singlet 
with respect to SO(4) with the index $i$ running from $1$ to 
$4$. In general, 
we can have similar singlet operators with dimensions $\Delta=J+2n
\,\,  (n=1, 2, \ldots)$. 
The authors of \cite{donos}  have derived a leg factor, 
which is a generalization of 
the kernel appearing in the LLM classification and is closely 
related to the one \cite{jeyo2} studied in connection with a 
deformation of  $c=1$ matrix model in which the deformation 
parameter may be related to the mass of 2D black hole. 
In view of mysteries associated with the leg factor 
in old matrix models, it would be 
desirable to further study this phenomenon from a 
new perspective of more general holographic correspondence 
between bulk and boundary theories.  

\vspace{0.2cm}
\noindent (6)\, \,  Finally, there are other important questions such as the 
study of possible M-theory version \cite{LLM}\cite{LM} 
 (or the case of AdS$_{4,7}\times$ S$^{7,4}$) of the present construction 
and general (non-conformal) D$p$-branes,\footnote{
In connection with this question for the case $p=0$, the model in 
\cite{jabbari} where 
 1/2-BPS states of D3-branes are 
treated in a framework of DLCQ matrix theories seems to be 
interesting.  
} 
formulation of T- and S-dualities, 
the deformation of backgrounds, and so on. 
In D-brane field theory, all possible background deformations  
would {\it in principle} be represented by fermion bilinear products, 
so that the notorious problem of treating RR-background 
fields may be viewed from a new perspective. 
It has also been suggested \cite{yo2} in the case of D-instantons that
 a full-fledged second quantized theory of D-branes 
may be relevant for approaching the question 
of background independent formulation.

\vspace{0.5cm}
\noindent
Acknowledgements

I would  like to thank A. Jevicki for stimulating conversations 
related to this subject at an early stage of the present work. 

A preliminary account of the present work was first 
given in the Workshop ``Cosmological Landscape: Strings, 
Gravity and Inflation", September 20-24, at KIAS, Seoul. 
I would like to thank the organizers for invitation and participants 
for their interests. 

The present work is supported in part by Grant-in-Aid for Scientific Research (No. 13135205 (Priority Areas) and No. 16340067 (B))  from the Ministry of  Education, Science and Culture, and also by Japan-US Bilateral Joint Research Projects  from JSPS.

\vspace{0.3cm}
\appendix 
\section{A derivation of free fermions for 
$\Delta=J$ operators}
\setcounter{equation}{0}
\renewcommand{\theequation}{\Alph{section}.\arabic{equation}}
\renewcommand{\thesubsection}{\Alph{section}.\arabic{subsection}}

The Hilbert space of the complex 1-matrix quantum mechanics 
is reducible to a system of $N$ free fermions. Although this has been 
discussed in the literature \cite{corley}\cite{takatsuchiya}, 
we give our own derivation here 
to set up our notations used in the main text. 
As discussed in the text, 
we think that the essence of the emergence of the 
fermion picture is actually 
not in the choice of a special two dimensional 
plane, but in the separation of degrees of freedom 
 into the Hermitian 1 matrix model and 
the kinematical degrees of SO(6). Use of the coherent state 
representation for the matrix part, however, leads to the 
completely equivalent formulation as the treatment of complex case
 given here. 

The (Euclidean) Hamiltonian is 
\EQ
H=\Tr\Big[
ZZ^{\dagger}+\Pi\Pi^{\dagger}
\Big]
\EN
where $Z, Z^{\dagger}$ and $\Pi, \Pi^{\dagger}$ are 
matrix-valued canonical coordinates and momenta 
satisfying the canonical commuation relations
\EQ
[Z_{nm}, \Pi_{m'n'}]=i\delta_{nn'}\delta_{mm'}, 
\quad [Z^{\dagger}_{nm}, \Pi^{\dagger}_{m'n'}]
=i\delta_{nn'}\delta_{mm'}
\EN
with all other commuators  vanishing. 
We use the notation $\dagger$ for matrix conjugartion and 
$*$ for the quantum-mechanical hermitian conjugation:
$Z_{mn}^{\dagger}=Z^*_{nm}, \, 
\Pi_{mn}^{\dagger}=\Pi^*_{nm}$ and 
also that in the $Z, Z^{\dagger}$-diagonal representation 
\EQ
\Pi_{mn}=-i {\partial\over \partial Z_{nm}}, 
\quad \Pi_{mn}^{\dagger}=-i{\partial \over \partial
Z^{\dagger}_{nm}}. 
\EN
In this representation, $Z^*_{nm}$ is of course the complex conjugate 
of $Z_{nm}$.  
We then define two sets of matrix-valued 
creation and annihilation operators by 
\EQ
A_{nm}\equiv 
{1\over \sqrt{2}}
(Z_{nm}+i\Pi^{\dagger}_{mn}), \quad 
A^{\dagger}_{mn}\equiv 
{1\over \sqrt{2}}
(Z_{nm}^{\dagger}-i\Pi_{mn}), 
\EN
\EQ
B_{nm}\equiv 
{1\over \sqrt{2}}
(Z_{mn}^{\dagger}+i\Pi_{nm}), \quad 
B^{\dagger}_{mn}\equiv 
{1\over \sqrt{2}}
(Z_{mn}-i\Pi^{\dagger}_{nm}). 
\EN
Only nonvanishing commutators among these 
oscillators are 
\EQ
[A_{nm}, A^{\dagger}_{m'n'}]=
\delta_{nn'}\delta_{mm'}=
[B_{nm}, B^{\dagger}_{m'n'}]
\EN
in terms of which the Hamiltonian operator is given as
\EQ
H=\Tr\Big[
A^{\dagger}A+B^{\dagger}B+N
\Big].
\EN
Since the 1/2-BPS operators satisfying the 
condition $\Delta=J$ consist of only the 
holomorphic coordinate $Z$, we can impose the following 
condition for allowed states:
\EQ
A_{nm}|\Psi\rangle=\Big(Z_{nm}+{\partial \over \partial Z^*_{nm}}
\Big) |\Psi\rangle =0
\EN
which in the $Z,Z^{\dagger}$ 
diagonal representation leads to 
\EQ
\langle Z, Z^{\dagger}|\Psi\rangle=
F[Z]\, \e^{-\Tr(ZZ^{\dagger})}
\EN
with $F[Z]$ being an arbitrary U($N$)-invariant 
holomorphic function 
of the matrix coordinate $Z$. Note that this is 
nothing but the general form of states in the 
{\it coherent-state representation} of the {\it hermitian} matrix model. 
The action of the Hamiltonian is 
\EQA
\langle Z, Z^{\dagger}|H|\Psi\rangle
&&=\Tr\Big[
-{\partial \over \partial Z^{\dagger}}{\partial \over \partial Z}+Z^{\dagger}Z
\Big]F[Z]\e^{-\Tr(ZZ^{\dagger})}\nonumber \\
&&=
(HF)[Z]\, \e^{-\Tr(ZZ^{\dagger})}
\EQN
where
\EQ
(HF)[Z]=\Big(\sum_{m, n}
Z_{nm}{\partial \over Z_{nm}} +N^2\Big)
F[Z]. 
\EN
The holomorphic functions $F[Z]$ are identified one-to-one with the 
operators ${\cal O}^J$. 
Thus apart from the zero-point contribution $N^2$, 
the eigenvalue of the Hamiltonian is equal to the 
conformal dimension. 

The norm of the holomorphic state is 
defined by the following internal product
\EQ
\langle \Psi_1|\Psi_2\rangle =
\int [dZdZ^{\dagger}]\langle \Psi_1|Z, Z^{\dagger}\rangle 
\langle Z, Z^{\dagger}|\Psi_2\rangle
=\int [dZdZ^{\dagger}]\e^{-2\Tr(ZZ^{\dagger})}
F_1[Z^{\dagger}]F_2[Z]. 
\EN
The conjugate 
operators ${\cal \overline{O}}^J$ are naturally 
identified with the anti-holomorphic function $F[Z^{\dagger}]$. 
It is easy to check that 
the action of the Hamiltonian is self-conjugate in the sense 
that 
\EQA
\langle \Psi_1|H|\Psi_2\rangle
&&=\int [dZdZ^{\dagger}]\e^{-2\Tr(ZZ^{\dagger})}
F_1[Z^{\dagger}](HF_2)[Z]
\nonumber \\
&&=
\int [dZdZ^{\dagger}]\e^{-2\Tr(ZZ^{\dagger})}
(HF_1)[Z^{\dagger}]F_2[Z]
\EQN
with
\EQ
(HF)[Z]^{\dagger}=\Big(\sum_{m, n}
Z_{nm}^*{\partial \over Z_{nm}^*} +N^2\Big)
F[Z^{\dagger}].
\EN

Since we are interested in the gauge invariant  
functions for $F[Z]$, it is natural to diagonalize the matrix 
coordinates. The symmetry which preserves both the 
Hamiltonian and the general states of the above form is the 
group of unitary transformations, 
\[
Z \rightarrow UZU^{\dagger}, \quad 
Z^{\dagger}\rightarrow UZ^{\dagger}U^{\dagger}, 
\quad UU^{\dagger}=1.
\]
The measure defined above for the internal product of holomorphic 
states is identical with the old standard form for the ensemble of 
random complex matrices, known as the 
Ginibre ensemble \cite{ginibre}. Though it is not possible to diagonalize a generic 
complex matrix by a unitary transformation, 
it is always possible to 
make it to the triangular form
\EQ
Z=U(\Lambda +K)U^{\dagger}, 
\EN
where $\Lambda$ is the complex diagonal matrix $\Lambda_{mn}=\delta_{mn}z_n$ with $z_n$'s being the eigenvalues of $Z$ and 
$K$ is a lower-triangular matrix satisfying $K_{mn}=0$ for  
$m\ge n$ (no diagonal elements). For arbitrary (anti-)holomorphic polynomials $F[Z]$ ($F[Z^{\dagger}]$), 
this is sufficient for  expressing the traces in terms of 
 eigenvalues $z_n$ ($z^*_n$). 
\[
\Tr(Z^k)=\sum_n z^k_n.
\] 
Up to a numerial factor, the integration measure is then given by
\EQ
[dZdZ^{\dagger}] \e^{-2\Tr(ZZ^{\dagger})}=[dU][dKdK^{\dagger}]
\Big(\prod_n dz_ndz^*_n\Big)
\Delta[z]\Delta[z^*] \exp\Big[-2\sum_n|z_n|^2 -
2\Tr(KK^{\dagger})\Big], 
\EN
where $[dU]$ is the invariant measure for the U($N$) group and 
\EQ
\Delta [z] =\prod_{n<m}(z_n-z_m) .
\EN
For the (anti-)holomorphic trace operators, the degrees of freedom of the lower-triangular part $K$  are 
decoupled and  can be integrated out giving 
a numerical factor. 
We warn the reader,  however,  that if we try to extend our 
treatment of correlators to non-extremal case, we have to 
include both holomorphic and anti-holomorphic 
variables into a single trace. Then, the decoupling 
of the triangular matrices $K$ is no more valid. Hence, 
the treatment of normal ordering prescription becomes 
more complicated as mentioned in the text. 

Thus, we are led to define the Hilbert space of multi-particle 
wave functions of the form 
\EQ
\langle z, z^*|\Psi\rangle =f[z]\, \e^{-2\sum_n |z|^2}, \quad 
f[z]= \Delta[z]F[z], 
\EN
which is sufficient for discussing extremal correlators. 
The internal product is 
\EQ
\langle \Psi_1|\Psi_2\rangle 
=\int \Big(\prod_n dz_ndz_n^*\Big)
f_1[z]^{*} f_2[z]\, \e^{-2\sum_n|z_n |^2}. 
\EN
The Hamiltonian acting on the holomorphic function $f[z]$ is 
\EQ
(hf)[z]=\Delta[z](HF)[Z]=(\Delta H\Delta^{-1}f)[z]. 
\EN
For conjugate states, all the quantities are replaced by 
their complex conjugates. 
We find 
\EQ
h=\Delta[z]\sum_n z_n{\partial \over \partial z_n}\Delta[z]^{-1} 
+N^2=
\sum_n z_n {\partial \over \partial z_n}+{1\over 2}N(N+1). 
\EN
Since $f[z]$ is completely antisymmetric with respect to the 
exchange of the eigenvalues, the system is equivalent with the 
system of $N$ free fermions. The set $\{f[z]\}$ of antisymmetric 
holomorphic functions  has a one-to-one 
correspondence to the set $\{F[Z]\}$ of the products of 
traces of the complex matrix $Z$. 
Aside from the constant 
contribution $N(N+1)/2$, the single-particle Hamiltonian 
is simply 
\EQ
h_s=z{\partial \over \partial z}
\EN
whose eigenfunctions are just the monomials $z^n$ with 
eigenvalue $n$. Therefore the ground-state energy of the 
$N$ fermions is equal to 
$\sum_{n=0}^{N-1}1=N(N-1)/2$ which gives the correct zero-point energy 
$N^2$ combined with the constant contribution 
$N(N+1)/2$. It is well known that this system 
is formally  identical with the one which emerges 
in describing the (integer) quantum Hall system in two 
dimensions.

\end{document}